\documentclass[twocolumn, 10pt, aps, pra, superscriptaddress,groupedaddress]{revtex4}
 
\usepackage{graphicx}
\usepackage{amsmath}
\usepackage{amssymb}
\usepackage{bm}
\usepackage{amsfonts}
\usepackage[usenames, dvipsnames]{xcolor}
\usepackage{appendix}
\usepackage[colorlinks=true, pdfstartview=FitV, linkcolor=blue, citecolor=blue, urlcolor=blue]{hyperref}
\usepackage{color}
\usepackage{epstopdf}
\usepackage{dsfont}
\graphicspath{{figures/}}

\renewcommand{\Re}{\operatorname{Re}}

\begin{document}
 
\title{Superfluid--Mott insulator transition of ultracold superradiant bosons in a cavity}
 
\author{Rui Lin}
\affiliation{Institute for Theoretical Physics, ETH Zurich, 8093 Zurich, Switzerland}
\author{Luca Papariello}
\affiliation{Institute for Theoretical Physics, ETH Zurich, 8093 Zurich, Switzerland}
\author{Paolo Molignini}
\affiliation{Institute for Theoretical Physics, ETH Zurich, 8093 Zurich, Switzerland}
\author{R. Chitra}
\affiliation{Institute for Theoretical Physics, ETH Zurich, 8093 Zurich, Switzerland}
\author{Axel U. J. Lode}
\affiliation{Wolfgang Pauli Institute c/o Faculty of Mathematics,
University of Vienna, Oskar-Morgenstern-Platz 1, 1090 Vienna, Austria}
\affiliation{Vienna Center for Quantum Science and Technology,
Atominstitut, TU Wien, Stadionallee 2, 1020 Vienna, Austria}

%%%%%%%%%%%%%%%%%%%%%%%%%%%%%%%%%%%%%-ABSTRACT-%%%%%%%%%%%%%%%%%%%%%%%%%%%%%%%%%%%%%
 
\begin{abstract}

We investigate harmonically-trapped, laser-pumped bosons with infinite-range interactions induced by a dissipative high-finesse red-detuned optical cavity with numerical and analytical methods.
We obtain multiple cavity and atomic observables as well as the full phase diagram of the system using the multiconfigurational time-dependent Hartree method for indistinguishable particles (MCTDH-X) approach. Besides the transition from an unorganized normal phase to a superradiant phase where atoms self-organize, we focus on an in-depth investigation of the self-organized superfluid to self-organized Mott insulator phase transition in the superradiant phase as a function of the cavity-atom coupling. The numerical results are substantiated by an analytical study of an effective Bose-Hubbard model.
We numerically analyze cavity fluctuations and emergent strong correlations between atoms in the many-body state across the Mott transition via the atomic density distributions and Glauber correlation functions.
Unexpectedly, the weak harmonic trap leads to features like a lattice switching between the two symmetry-broken $\mathbb{Z}_2$ configurations of the untrapped system and a reentrance of superfluidity in the Mott insulating phase. Our analytical considerations {\it quantitatively} explain the numerically observed correlation features.

\end{abstract}

\maketitle

%%%%%%%%%%%%%%%%%%%%%%%%%%%%%%%%%%%%%%-INTRO-%%%%%%%%%%%%%%%%%%%%%%%%%%%%%%%%%%%%%%

\section{Introduction}

Enormous progress in preparing and controlling ultracold atoms in high-finesse optical cavities has been instrumental in bringing the strongly coupled regime of cavity quantum electrodynamics to experimental reality~\cite{KimbleQED, BrenneckeQED, ColombeQED}.
In particular, Bose-Einstein condensates (BEC) in optical cavities~\cite{Ritter2009, PRLPurdy} have emerged as a successful platform to study novel phases of matter stabilized by light.
The strong light-matter coupling achieved in these systems of light and matter has led to the realization of the long-sought Hepp-Lieb superradiant phase transition in the Dicke model~\cite{dicke54, hepp73, wang73, carmichael73, Nagy2010, baumann10}.  
Cavity-matter coupled systems have since then also provided a versatile arena to realize novel topological states~\cite{deng14, dong14,mivehvar17}, limit cycles~\cite{piazza15} and various correlated phases in the presence of an optical lattice~\cite{niederle16,dogra16,landig16, hruby18}, with recent experiments on spinor condensates highlighting even the formation of spin textures~\cite{landini18}.

Cavity engineering thus provides a promising path to devising new exotic correlated phases of matter.
The self-organized superradiant state in a cavity-BEC system offers an excellent example of a light-matter phase in which correlations play a crucial role. 
At a critical atom-field coupling, a quantum phase transition from a \emph{normal phase} (NP) to a \emph{superradiant phase} (SP) manifests itself as self-organization of the atoms of the condensate into a lattice~\cite{nagy08,baumann10}.
In the SP, a further transition between an uncorrelated \emph{self-organized superfluid} (SSF) phase and a correlated \emph{self-organized Mott insulator} (SMI) phase can be observed at higher pump powers~\cite{klinder15}.
This cavity-induced transition is fundamentally similar to the Mott transition observed in Bose-Hubbard models realized in optical lattices~\cite{greiner02, kato08, tomita17}.
The SSF-SMI transition cannot be captured with standard Gross-Pitaevskii mean-field theories usually used to describe the atomic ensemble in such systems, because these theories neglect the correlations present in the SMI phase~\cite{axel17}.
The Mott transition was studied qualitatively by mapping the system to an effective Bose-Hubbard model in Ref.~\cite{bakhtiari15,panas17}.
However, a full examination of the transition in this problem --- including the role of cavity fluctuations and associated many-body correlations --- is still lacking.

A quantitative study of the emergence and properties of strongly correlated phases in these systems is rather complex.
On the one hand, typical theoretical approaches rely on mean-field techniques~\cite{maschler08,nagy08,Nagy2010}, which are tailored to capturing only a particular phase and ignore quantum fluctuations and correlations.
On the other hand, most of the numerical methods designed to study correlated phases are adapted to lattice systems, whereas cavity-matter systems may, depending on the pumping strength, require a continuum-description.
An ideal methodology for cavity-BEC systems should be able to deal with both the continuum and discrete spatial aspects of the problem at the same degree of complexity.
A tool that enables such a numerical description of correlated many-body states with quantum fluctuations is the Multiconfigurational Time-Dependent Hartree method for indistinguishable particles (MCTDH-X)~\cite{ultracold,axel16,alon08,fasshauer16}. MCTDH-X uses a variationally optimized basis set to represent the many-body wave function for realistic experimental setups.

In the present work, we use this methodology to study the full phase diagram of a single-component BEC coupled to a red-detuned cavity in one spatial dimension.
We rigorously study the standard NP-SP transition as well as the SSF-SMI transition in the self-organized phase. 
In previous studies of the problem using MCTDH-X~\cite{axel17,axel18}, cavity fluctuations were structurally ignored, and an in-depth comprehension of the origin of the Mott-insulator phase and the emergence of so-called ``fragmentation'' in this phase was lacking. \emph{Fragmentation} is present in a many-body system of bosons if the reduced one-body density matrix of the quantum state has more than one macroscopic eigenvalue\cite{spekkens99,mueller06}. Fragmentation is a many-body effect intertwined with the emergence of quantum correlations and fluctuations, and thus it cannot be captured by mean-field approaches\cite{sakmann08}. In this work, we go beyond the mean-field approach in our description of both the cavity and the atomic fields, and demonstrate how cavity fluctuations build up close to the NP-SP transition and how strong correlations between the atoms' positions and momenta emerge.
We find that the presence of the harmonic trap leads to an interesting switching between the two $\mathbb{Z}_2$ symmetry-broken configurations in the superradiant phase and a reentrance of superfluid features in the Mott phase. Both phenomena stem from the competition between the harmonic trapping potential and the particle-particle interaction energy.
Additionally, we provide a mapping of the cavity-BEC system to a Bose-Hubbard (BH) model, whose results are in mutual agreement with the numerical ones obtained using MCTDH-X.

This paper is structured as follows. 
In Section ~\ref{sec:model}, we introduce the cavity-BEC system and show that the dissipative cavity leads to an effective Hamiltonian for the bosonic atoms with infinite-range two-body interactions that retains the non-mean-field properties of both the cavity field and the atomic field. 
Our numerical and analytical results on the phase diagram and phase transitions of the cavity-BEC system are presented in Section~\ref{sec:results}. In the end, we conclude in Section~\ref{sec:conclusion}.

%%%%%%%%%%%%%%%%%%%%%%%%%%%%%%%%%%%%%-SECTION 1-%%%%%%%%%%%%%%%%%%%%%%%%%%%%%%%%%%%%%

\section{Model, method and quantities of interest}\label{sec:model}

\subsection{Model}

\begin{figure}[t]
	\centering
	\includegraphics[width=0.9\columnwidth]{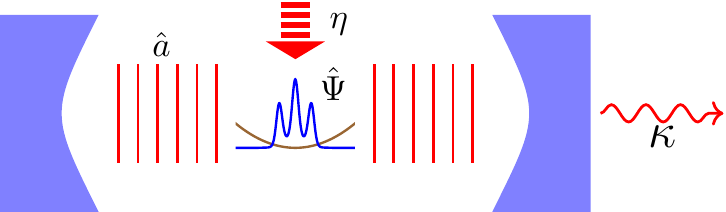}
	\caption{Schematic setup of a harmonically-trapped driven BEC in a dissipative optical cavity. }
	\label{schematics}
\end{figure}

We study a BEC of $N$ weakly-interacting and harmonically-trapped ultracold bosons coupled to a high-finesse lossy optical cavity. The cavity has a single mode of frequency 
$\omega_c$ and wave vector $k_c$. A pump laser of frequency $\omega_p$ coherently drives the BEC perpendicular to the cavity axis. 
We study a one-dimensional version of the system as sketched in Fig.~\ref{schematics}. As we will show later, the cavity leads to infinite-range interactions, implying that the
actual spatial dimensionality does not affect the physics of the system qualitatively. The one-dimensional cavity-BEC system can be described by the following Hamiltonian in the
rotating frame ~\cite{maschler08},
\begin{align} \label{hamil}
\hat{\mathcal{H}} & = \int \mathrm{d}x \hat{\Psi}^\dagger(x)\left\{\frac{p^2}{2m}+\frac{g}{2}\hat{\Psi}^\dagger(x)\hat{\Psi}(x)+V_\text{trap}(x)\right\}\hat{\Psi}(x) 
\nonumber \\
& \quad +\hbar U_0\hat{a}^\dagger\hat{a}\int \mathrm{d}x \hat{\Psi}^\dagger(x)\hat{\Psi}(x) \cos^2(k_c x) \nonumber \\
& \quad +\hbar\eta(\hat{a}+\hat{a}^\dagger)\int \mathrm{d}x \hat{\Psi}^\dagger(x)\hat{\Psi}(x) \cos(k_c x) \nonumber \\
& \quad -\hbar\Delta_c\hat{a}^\dagger\hat{a} \, .
\end{align}
Here, $\hat{\Psi}(x)^{(\dagger)}$ denote atomic field operators at position $x$ and $\hat{a}^{(\dagger)}$ represent photonic operators.
The atoms are confined by an external harmonic trap $V_\text{trap}(x)=\frac{1}{2}m\omega_x^2x^2$ and interact repulsively through a contact interaction with 
strength $g$. The parameter $U_0$ encodes the light shift of a single maximally-coupled atom, while $\eta$ is the two-photon Rabi frequency describing the fluctuations of the light-atom interaction. In the following, we refer to $\eta$ as pump rate for convenience. 
Finally, $\Delta_c = \omega_p - \omega_c\ll \omega_p,\,\omega_c$ is the detuning between the pump frequency and the cavity resonance frequency. 
The lossy cavity is characterized by a dissipative rate given by $\kappa$.

This system has been amply studied within a mean-field approximation where the cavity field operator is substituted by its expectation value $\alpha=\langle \hat{a}\rangle$ and the atoms are treated within the Gross-Pitaevskii mean-field approach~\cite{nagy08,Nagy2010,piazza15}. Under this approximation, the cavity-atom coupling generates an effective one-body potential for the bosons, as shown in detail in Appendix~\ref{sec:OBP}. Therefore, we call this the \emph{effective potential} approach. However, the mean-field treatment of the cavity field ignores the cavity and atomic field fluctuation around the NP-SP critical point. Additionally, the Gross-Pitaevskii mean-field treatment of the atomic field makes it impossible to capture the experimentally observed SMI phase~\cite{klinder15}, which is a correlated phase of matter. 

To access both the fluctuations at the critical point and the SMI phase, one needs to treat fluctuations in a self-consistent manner which requires the study of the full dynamical behavior of the operators.
Since the cavity is dissipative, assuming a Lindblad approach to dissipative systems~\cite{breue02}, the cavity field is described by the following equation of motion~\cite{maschler08}
\begin{eqnarray}
\frac{\partial}{\partial t}\hat{a}&=&\left[i\Delta_c-iU_0
\hat{\mathcal{B}}-\kappa\right]\hat{a}-i\eta\hat{\Theta},
\end{eqnarray}
with
\begin{subequations}\label{eq:b_and_theta}
\begin{eqnarray}
\hat{\mathcal{B}}&=&\int \mathrm{d}x \hat{\Psi}^\dagger(x)\hat{\Psi}(x) \cos^2(k_c x)\\
\hat{\Theta}&=&\int \mathrm{d}x \hat{\Psi}^\dagger(x)\hat{\Psi}(x) \cos(k_c x).\label{def_theta}
\end{eqnarray}
\end{subequations}
Since we are interested in the steady-state physics of the system [Eq.~\eqref{hamil}], we first adiabatically eliminate the  cavity field by setting $\partial_t \hat{a}=0$. This simplification is valid  for lossy cavities for which $\hbar\kappa\gg\hbar^2k_c^2/2m$, where the cavity adiabatically follows the atomic motion~\cite{maschler08}. 
The steady-state solution of the cavity field operator is formally given by~\cite{maschler08}
\begin{eqnarray}\label{eq:photon_steady}
\hat{a}=\eta [\Delta_c-U_0\hat{\mathcal{B}}+i\kappa]^{-1}\hat{\Theta}.
\end{eqnarray}

Inserting Eq.~\eqref{eq:photon_steady} in the Hamiltonian Eq.~\eqref{hamil}, we notice that the operator $\hat{\mathcal{B}}$ always appears in the form $\Delta_c-U_0\hat{\mathcal{B}}$, including in the second and fourth lines of Eq.~\eqref{hamil}. In the limit $|\Delta_c|\gg |NU_0|$, the density profile is mainly determined by the operator $\hat{\Theta}$, while the operator $\hat{\mathcal{B}}$ can be approximated by its expectation value, the bunching parameter $\mathcal{B}=\langle\hat{\mathcal{B}}\rangle$~\cite{maschler08}. Consequently, we obtain the following Hamiltonian for the atoms~\cite{mottl14}
\begin{eqnarray} \label{eq:LRI_appr}
\hat{\mathcal{H}} & =& \int \mathrm{d}x \hat{\Psi}^\dagger(x)\left\{\frac{p^2}{2m}+\frac{g}{2}\hat{\Psi}^\dagger(x)\hat{\Psi}(x)+V_\text{trap}(x)\right\}\hat{\Psi}(x) 
\nonumber \\
&&+\frac{\hbar\eta^2(\Delta_c-\mathcal{B}U_0)}{(\Delta_c-\mathcal{B}U_0)^2+\kappa^2}\times\nonumber\\
&&\int \mathrm{d}x \mathrm{d}x' \hat{\Psi}^\dagger(x)\hat{\Psi}^\dagger(x')  \cos(k_cx)\cos(k_cx')\hat{\Psi}(x)\hat{\Psi}(x'). \nonumber\\
\end{eqnarray}
The cavity therefore effectively induces an infinite-range two-body interaction between the atoms, as described by the third line of Eq.~\eqref{eq:LRI_appr}. For this reason, we call it the \emph{long-range interaction} (LRI) approach.
Note that if $\hat{\mathcal{B}}$ is retained as an operator, three- and many-body interactions among the atoms occur, because $\hat{\mathcal{B}}$ is a function of the atomic field operator [Eq.~\eqref{def_theta}].
To understand the steady-state properties of the system, it suffices to investigate the ground state of the LRI Hamiltonian [Eq.~\eqref{eq:LRI_appr}]. Deep in the red-detuned cavity region $-\Delta_c>|NU_0|$~\cite{baumann10,klinder15}, the prefactor of this interaction term is negative.
In this case, the induced two-body interaction is minimized by two lattice configurations of the atoms. In the even lattice, the  atoms are localized at the maxima of the function $\cos(k_cx)$ [$x=2n\pi/k_c$, $n\in\mathbb{N}$], while in the odd lattice, they are localized at the minima [$x=(2n+1)\pi/k_c$, $n\in\mathbb{N}$]. This implies that the atoms of the condensate can in principle self-organize into a lattice with spacing $2\pi/k_c$.
In the absence of the harmonic trap, these are the two configurations of the spontaneously broken $\mathbb{Z}_2$ symmetry accompanying the normal phase (NP)- superradiant phase (SP) transition~\cite{baumann11}.

\subsection{Method}

To study the full effect of the long-range interactions, we analyze the problem using MCTDH-X~\cite{ultracold,axel16,alon08,fasshauer16}. This is a variational method used to obtain the many-body wave function of the system. 
The ansatz for the wave function
\begin{eqnarray}
|\Psi\rangle=\sum_{\mathbf{n}=\{n_1,...,n_M\}} C_\mathbf{n}(t)\prod^M_{k=1}\left[ \frac{(\hat{b}_k^\dagger(t))^{n_k}}{\sqrt{n_k!}}\right]|0\rangle,
\end{eqnarray}
is composed of a total number of $M$ adaptive single-particle wave functions, or \emph{orbitals} $\psi_i(x;t)$, each of which is created by an operator $\hat{b}_i^\dagger(t)$,
\begin{equation}
 \psi_i(x;t) = \langle x \vert \hat{b}_i^\dagger(t) \vert 0 \rangle.
\end{equation}
Since both the orbitals $\psi_i(x;t)$ and the coefficients $C_\mathbf{n}(t)$ vary in time, the adaptive single-particle basis enables MCTDH-X to optimally represent the many-body wave function of interacting bosons.
Using imaginary time propagation, the wave functions and energies of the low-lying steady states of the problem can be obtained.
MCTDH-X can capture the correlations between the bosons and study the effects of inhomogeneities like the harmonic trap. Details on the theory of MCTDH-X are presented in Appendix~\ref{sec:mctdhx}.

The concept of orbitals, as well as their total number $M$, are important for correctly capturing the correlations [see Appendix~\ref{sec:mctdhx}]. With a single orbital $M=1$, MCTDH-X is reduced to the Gross-Pitaevskii mean-field limit, where the many-body state $\Psi$ is described by a product of single-particle wave functions $\psi$, $\Psi(x_1,...,x_N)=\prod_{i=1}^N\psi(x_i)$. 
With $M>1$, MCTDH-X works beyond mean-field, and the many-body state is a superposition of many
symmetrized states of $N$ particles in at most $M$ different single-particle
wave functions. 
For a large orbital number $M$, MCTDH-X approaches the exact solution and all the correlations inside the system are captured~\cite{alon08,axel12,fasshauer16}.  Strongly localized phases like the Mott insulator may need many orbitals to correctly capture the physics~\cite{axel162}. For a given number of particles $N$, since the required computational resources scale as $\binom{N+M-1}{N}$~\cite{alon08}, the requirement of convergence with respect to the number of orbitals $M$ constrains the number of particles that can be studied. In our simulations, it suffices to work with $M=5$ orbitals to obtain converged results for a large range of particle numbers.

In the Gross-Pitaevskii mean-field limit, the Hamiltonian Eqs.~\eqref{hamil} and \eqref{eq:LRI_appr} are invariant under rescaling of the atom number $N \to N'$, provided that the parameters $g$, $U_0$ and $\eta$ change according to~\cite{nagy08}
\begin{eqnarray} \label{numberscale}
g\mapsto\frac{N}{N'}g,\quad
U_0\mapsto \frac{N}{N'} U_0,\quad \eta\mapsto \sqrt{\frac{N}{N'}}\eta .
\end{eqnarray}
It can be verified that the MCTDH-X results for $M=1$ orbital are consistent with this scaling. Beyond the Gross-Pitaevskii mean-field, this scaling is invalid. To obtain results like correlations for large number $N$ of atoms beyond $M=1$, we explicitly simulate systems with different atom numbers $N=50$, 100, and extrapolate the results to higher $N$. While changing the atom number $N$, we rescale parameters according to Eq.~\eqref{numberscale}. In this way, the mean-field properties are kept unchanged and it is easier to extract physics beyond mean-field from our computations.

\subsection{Quantities of interest}

The properties of the atoms, especially the correlations within the system, can be revealed by density distributions and correlation functions. These include the density distribution in position and momentum spaces $\rho(x)$ and $\tilde{\rho}(k)$
\begin{subequations}
	\begin{eqnarray}
	\rho(x)&=&\rho^{(1)}(x,x)/N,\\
	\tilde{\rho}(k)&=&\langle\hat{\Psi}^\dagger(k)\hat{\Psi}(k)\rangle/N,
	\end{eqnarray}
\end{subequations}
as well as the one-body and two-body Glauber correlation functions $g^{(1)}(x,x')$ and $g^{(2)}(x,x')$
\begin{subequations}\label{eq:def_g1g2}
	\begin{eqnarray}
	g^{(1)}(x,x')&=&\frac{\rho^{(1)}(x,x')}{\sqrt{\rho^{(1)}(x,x)\rho^{(1)}(x',x')}},\\
	g^{(2)}(x,x')&=&\frac{\rho^{(2)}(x,x')}{\rho^{(1)}(x,x)\rho^{(1)}(x',x')}, \label{eq:def_g2}
	\end{eqnarray}
\end{subequations}
where $\rho^{(1)}(x,x')$, $\rho^{(2)}(x,x')$ are the one-body and two-body reduced density matrices in position space, which are defined as
\begin{subequations}\label{eq:def_rho}
	\begin{eqnarray}
	\rho^{(1)}(x,x')&=&\langle\hat{\Psi}^\dagger(x)\hat{\Psi}(x')\rangle,\\
	\rho^{(2)}(x,x')&=&\langle\hat{\Psi}^\dagger(x)\hat{\Psi}^\dagger(x')\hat{\Psi}(x)\hat{\Psi}(x')\rangle.
	\end{eqnarray}
\end{subequations}
The Glauber correlation functions $g^{(1)}$ and $g^{(2)}$ quantify the local first- and second-order coherence of the many-body state.
First- (second-)order coherence is achieved when $|g^{(1)}|= 1$ ($g^{(2)} =1$) holds, and
implies that the reduced one- (two-)body density matrix $\rho^{(1)}$ ($\rho^{(2)}$) can be sufficiently described by one single-particle state. The proximity of $|g^{(1)}|$ and $g^{(2)}$ to unity gives a quantitative and spatially resolved measure for the departure of a many-body state from a Gross-Pitaevskii mean-field description, which presupposes that the many-body system is described by a product of one effective single-particle state~\cite{glauber63,sakmann08}.

Many physical phenomena, like the transition between phases, are reflected in the natural orbitals $\phi_i(x)$ and their occupancies $\rho_i$, which are defined, respectively, as the eigenvectors and eigenvalues of the one-body reduced density matrix $\rho^{(1)}$,
\begin{eqnarray}\label{eq:occupancy}
\rho^{(1)}(x,x') = \sum_{i=1}^M N \rho_i \phi^*_i(x';t) \phi_i(x;t),
\end{eqnarray}
with $\rho_1\ge\rho_2\ge...\ge\rho_M$. A state is said to be \emph{fragmented} if more than one orbital is macroscopically occupied, i.e., if $\rho_2$ becomes significant. 
Fragmentation is closely related to non-trivial correlations between atoms~\cite{nozieres95,sakmann08,spekkens99,penrose56}. For example, a BEC is non-fragmented, while a Mott insulator is highly-fragmented~\cite{sakmann08,roy18,chatterjee17}.

%%%%%%%%%%%%%%%%%%%%%%

%%%%%%%%%%%%%%%%%%%%%%%%%%%%%%%%%%%%%-SECTION 2-%%%%%%%%%%%%%%%%%%%%%%%%%%%%%%%%%%%%%

\section{Results}\label{sec:results}
We now describe the predictions of MCTDH-X for the cavity-BEC problem and the associated analytical mapping to a Bose-Hubbard model. The simulations use roughly the same experimental parameters for the $^{87}$Rb atom gas and the optical cavity as given in Ref.~\cite{baumann10}, which are listed in detail in Appendix~\ref{sec:parameters}. In principle, MCTDH-X is also applicable to simulate other experimental realizations. In this work, all the parameters and results of the simulations are given and presented in dimensionless units unless specified otherwise, see Appendix~\ref{sec:parameters}.

\begin{figure}[t]
	\centering
	\includegraphics[width=\columnwidth]{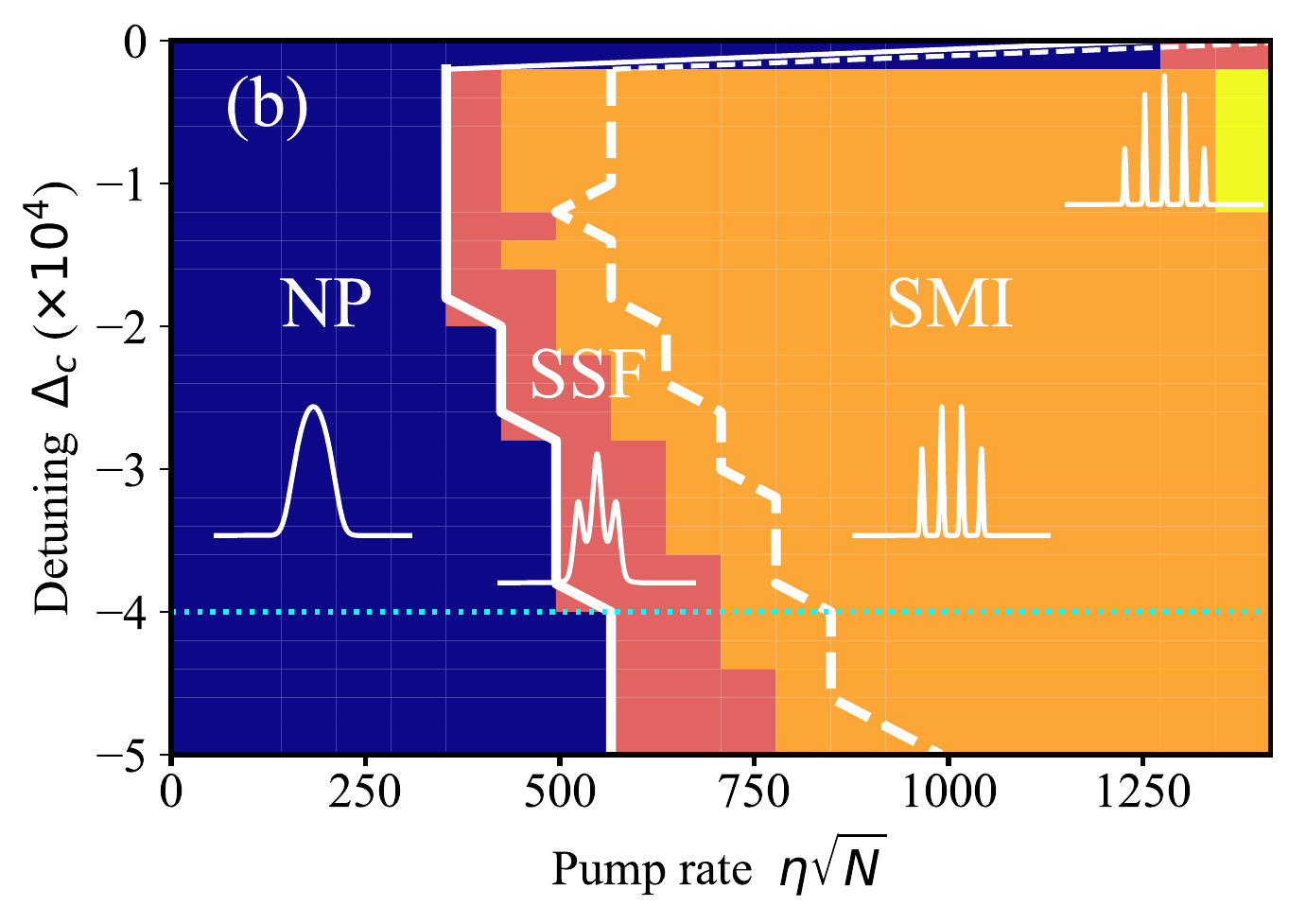}
	\caption{The full phase diagram of the BEC coupled to a cavity as a function of the dimensionless cavity detuning and the dimensionless pump rate. The phase diagram is divided into three phases: the normal phase (NP), the self-organized superfluid (SSF) phase and the self-organized Mott insulator (SMI) phase. The solid and dashed white lines indicate the transitions NP-SSF and SSF-SMI, respectively. 
	The colors correspond to the number of peaks in the spatial density profile, where blue, red, orange, and yellow correspond to NP, 3, 4 and 5 peaks in the position space density $\rho(x)$, respectively. All the other parameters used in these simulations and their dimensionless counterparts are described in Appendix \ref{sec:parameters}.
	The results discussed in the rest of this work are obtained by fixing the detuning at $\Delta_c=-4\times10^4$ ($2\pi\times10.1$MHz) and varying the pump rate along the dotted light blue line.}
	\label{diagram}
\end{figure}

\begin{figure*}[!t]
	\centering
	\includegraphics[width=\textwidth]{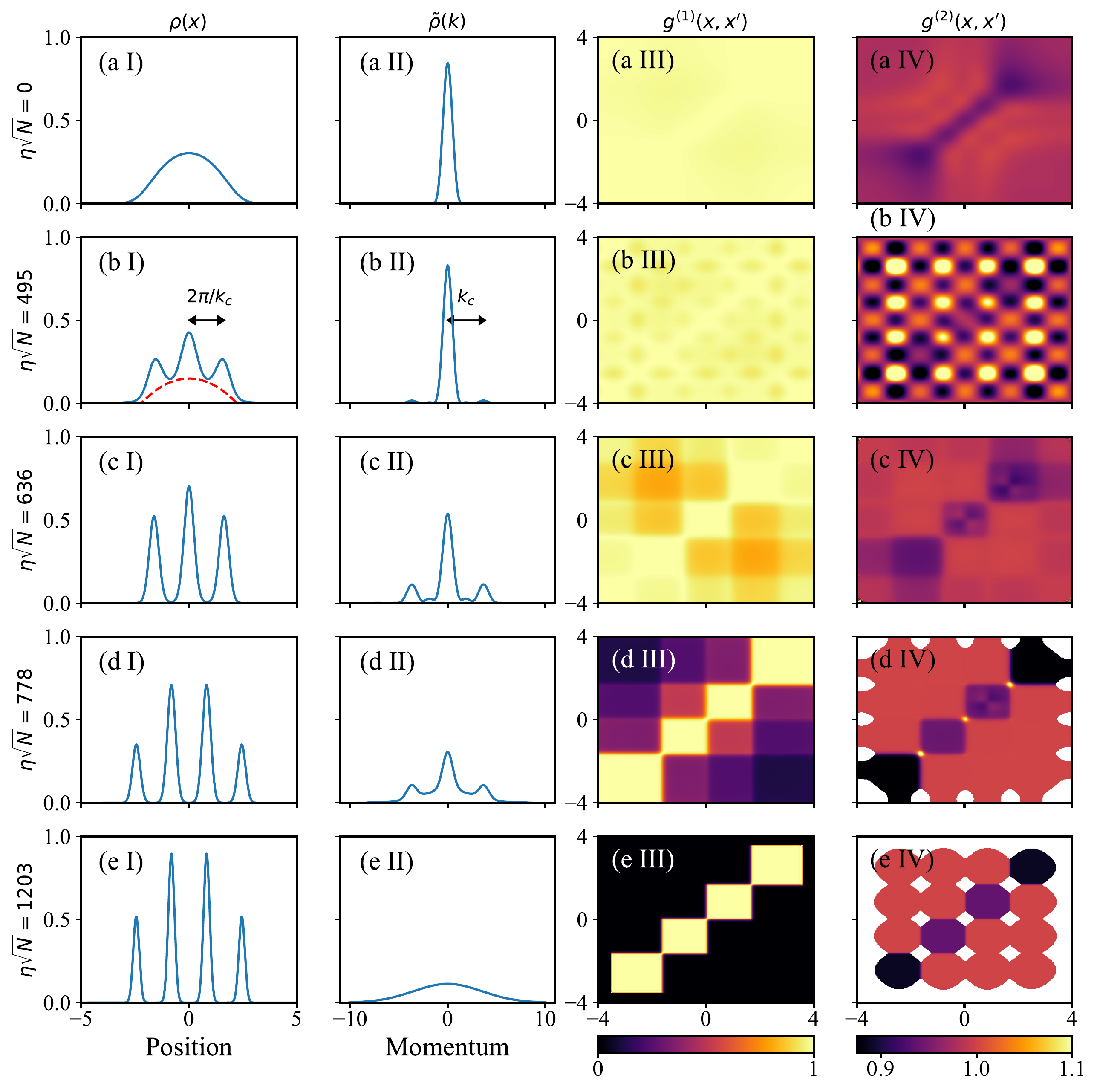}
	\caption{
		Density distributions and correlation functions in different phases as functions of pump rate at fixed cavity detuning $\Delta_c=-4\times10^4$ ($2\pi\times10.1$MHz), corresponding to the light blue line in Fig.~\ref{diagram}. The vertical panel (I) corresponds to the position space density distributions $\rho(x)$, panel (II) the momentum space density distributions $\tilde{\rho}(k)$, panel (III) the one-body correlation functions $\rho^{(1)}(x,x')$, and panel (IV) the two-body correlation functions $\rho^{(2)}(x,x')$, respectively. Each horizontal panel corresponds to a given pump rate. The pump rates have been chosen such that panel (a) describes the physics deep in a NP state, where the BEC has a Thomas-Fermi spatial profile and trivial correlations. Panel (b) shows a SSF state near the NP-SSF boundary, where the system starts to self-organize and the critical behaviors like cavity fluctuations and atomic long-range correlations are significant. The red dashed parabola in (b I) indicates the superfluid portion. Panel (c) features a SSF state far from both transition boundaries, where in momentum space the peaks corresponding to spatial modulations become significant, while the cavity fluctuation becomes negligible again. Panel (d) depicts a state near the SSF-SMI boundary, where non-local correlations between atoms start to decrease. Lastly, panel (e) portrays a SMI state, where the atoms are completely locally correlated.
		The white spaces in (d IV) and (e IV) come from the fact that we only plot $g^{(2)}(x,x')$ where $\rho^{(1)}(x,x)\rho^{(1)}(x',x')>10^{-7}$ to avoid numerical singularities. }
	\label{densities}
\end{figure*}
\clearpage

One key result of our simulations is the phase diagram in Fig.~\ref{diagram} and the corresponding density distributions and correlation functions in Fig.~\ref{densities}.
In the phase diagram, the transitions from a normal phase (NP) to a self-organized superfluid phase (SSF) to a self-organized Mott insulator phase (SMI) can be clearly seen. Additionally, the number of peaks in the position space profile increases from 3 to 4 to 5, implying configuration switchings between the even and odd lattices [cf. Figs.~\ref{densities}(b I)-(e I) and \ref{xexpansion}]. In Fig.~\ref{densities} and the rest of this work, the cavity detuning is fixed at $\Delta_c=-4\times 10^4$ ($2\pi\times10.1$MHz) in all simulations.
Fig.~\ref{densities} shows the density distribution in position space $\rho(x)$ and momentum space $\tilde{\rho}(k)$, as well as the one-body $g^{(1)}(x,x')$ and two-body $g^{(2)}(x,x')$ Glauber correlation functions for five states with different pump rates. The values of these pump rates correspond to a state in the NP, one on the boundary between NP and SSF, one in the SSF, on the boundary between SSF and SMI, and one in the SMI, respectively. The features of these simulation results will be discussed in detail in the following sections.

Part of the features shown in these two figures have already been discussed by Ref.~\cite{axel17}. A crucial difference in methodology between the work presented here and in Ref.~\cite{axel17} is that we use the LRI approach, treating the action of the cavity photons on the atoms as an infinite-range two-body interaction and retaining the cavity fluctuation. 
Furthermore, Ref.~\cite{axel17} focused on fragmentation of the condensate as a function of pump rate using the effective potential approach mentioned earlier [see Appendix~\ref{sec:OBP}]. We will show that the change in fragmentation is indeed related to the SSF-SMI transition.

\subsection{Normal - superradiant phase transition}\label{subsec:NP-SP}
As the pump rate $\eta$ increases, the cavity-BEC system is expected to transition from the normal phase (NP) to the superradiant phase (SP).
Figs.~\ref{densities} (a)-(c) show how the density distributions and correlation functions evolve from a NP state with $\eta\sqrt{N}=0$ to a SSF state with $\eta\sqrt{N}=636$ ($2\pi\times$160kHz). At a low pump rate, the system is in the NP. It has a Thomas-Fermi cloud profile in position space [Fig.~\ref{densities}(a I)] and correspondingly a high and narrow distribution centered at $k=0$ in momentum space [Fig.~\ref{densities}(a II)]. Its $g^{(1)}$ is trivially unity [Fig.~\ref{densities}(a III)], but its $g^{(2)}$ shows small structures differing slightly from $1$ [Fig.~\ref{densities}(a IV)]. These structures imply that the system is close to but slightly beyond a Gross-Pitaevskii mean-field description.
As the pump rate increases, the system goes into the SP, where the atoms self-organize into a lattice. Near the phase boundary, the atoms start to gather around the lattice sites, but a clear superfluid portion can still be seen in the density profile, as indicated by the red parabola in Fig.~\ref{densities}(b I). In momentum space, two peaks at $k=\pm k_c$ start to appear, corresponding to the periodic modulation in position space [Fig.~\ref{densities}(b II)]. In the correlation functions $\rho^{(1)}(x,x')$ and $\rho^{(2)}(x,x')$,  fluctuations with a period of $2\pi/k_c$ in both $x$ and $x'$ directions can be seen [Figs.~\ref{densities}(b III) and (b IV)]. These atomic field fluctuations are related to the cavity field fluctuations $\langle \delta\hat{a}^2\rangle$ discussed in the following paragraphs.
As the increasing pump power drives the system further into the SSF phase, the peaks in position space become separated from each other [Fig.~\ref{densities}(c I)]. In momentum space, the two peaks at $k=\pm k_c$ become more discernible while the height of the central peak $\tilde{\rho}(k=0)$ becomes significantly lower than before. In the correlation functions [Figs.~\ref{densities}(c III) and (c IV)], the fluctuations vanish and the superfluid features dominate again, with additional features stemming from the self-organization.

The  NP-SP transition can also be characterized by the cavity order parameter  $|\langle\hat{a}\rangle|$  
and the cavity fluctuations $\langle \delta\hat{a}^2\rangle= \langle  \hat{a}^\dagger \hat{a}\rangle-\langle  \hat{a}^\dagger\rangle\langle \hat{a}\rangle$ ~\cite{nagy08,brennecke13,nagy11}. Within the LRI approach, the cavity field operator [see Eq.~\eqref{eq:photon_steady}] can be approximated by 
$\hat{a}\approx \frac{\eta}{\Delta_c} \int \mathrm{d}x \hat{\Psi}^\dagger(x)\hat{\Psi}(x)\cos(k_cx)$  when $|\Delta_c|\gg NU_0,\kappa$. The expectation values of the cavity field and fluctuations can then be calculated by
\begin{subequations} \label{eqs:exp}
\begin{align}
|\langle\hat{a}\rangle| =& \left|\frac{\eta}{\Delta_c}\int \mathrm{d}x \rho^{(1)}(x,x)\cos(k_cx)\right|  \\
\langle \hat{a}^\dagger\hat{a}\rangle 
=&\frac{\eta^2}{\Delta_c^2}\int\rho^{(2)}(x_1,x_2)\cos(k_cx_1)\cos(k_cx_2) \mathrm{d}x_1 \mathrm{d}x_2 \nonumber \\
&+ \frac{\eta^2}{\Delta_c^2}\int \rho^{(1)}(x_1,x_1)\cos^2(k_c x_1) \mathrm{d}x_1,
\end{align}
\end{subequations}
where the reduced density matrices $\rho^{(1)}$ and $\rho^{(2)}$ are defined in Eq.~\eqref{eq:def_rho}.

\begin{figure}[t]
	\centering
	\includegraphics[width=\columnwidth]{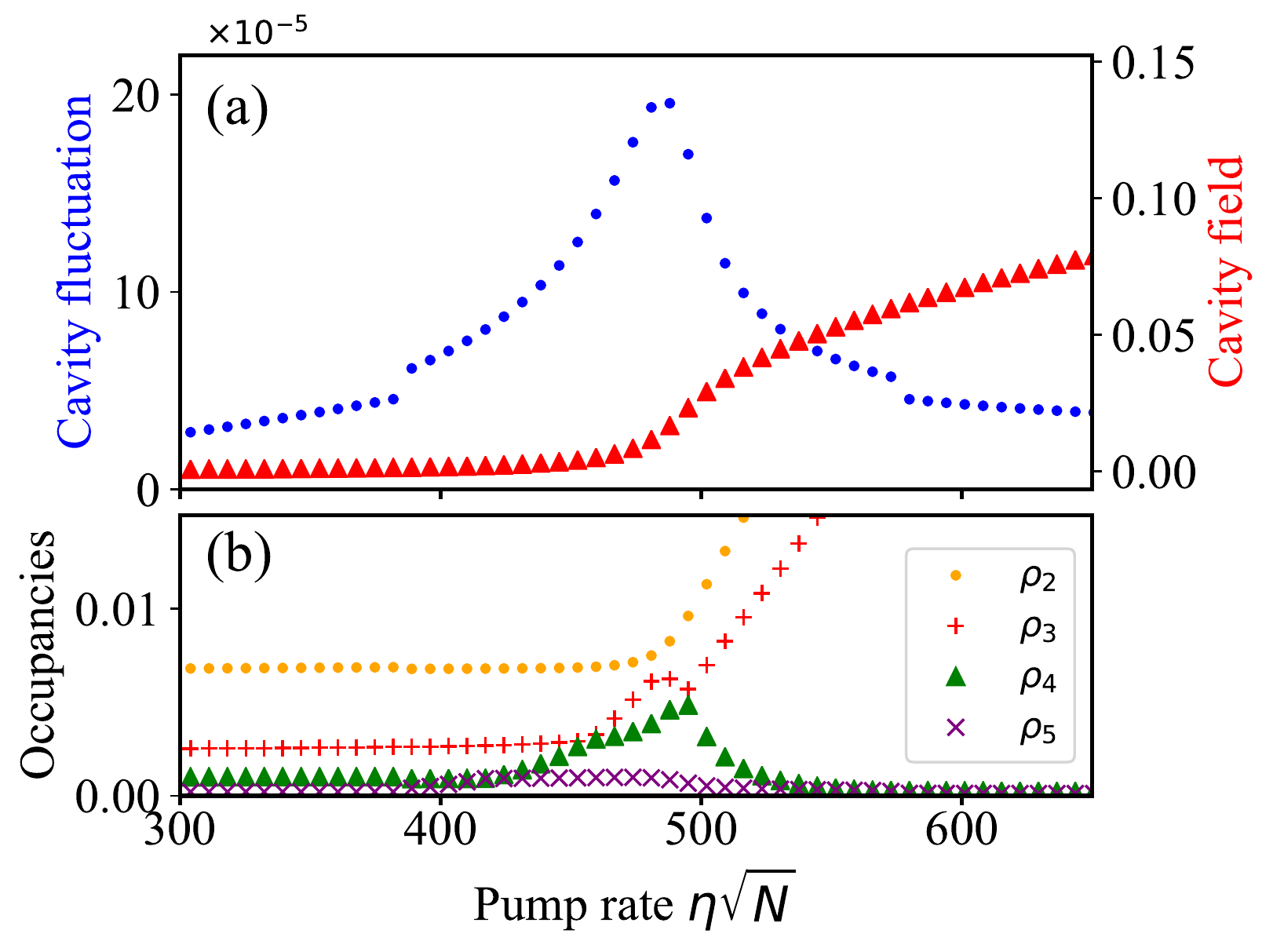}
	\caption{(a) The expectation values of the cavity field $|\langle\hat{a}\rangle|$ (red triangles, coordinates on right) and the cavity field fluctuation $
	\langle\hat{a}^\dagger\hat{a}\rangle-\langle\hat{a}^\dagger\rangle\langle\hat{a}\rangle$ (blue dots, coordinates on left) as functions of the pump rate $\eta\sqrt{N}$. The cavity fluctuations are strongly enhanced in the vicinity of the critical pump rate $\eta_c\sqrt{N}=480$ ($2\pi\times121$kHz). 
	(b) The orbital occupancies $\rho_i$ as functions of the pump rate $\eta\sqrt{N}$ [cf. Eq~\eqref{eq:occupancy}]. The first orbital is maximally occupied, so we only plot the occupancies of the higher order orbitals ($\rho_2$ to $\rho_5$ from top to bottom). Higher order orbitals are more significantly occupied in the vicinity of the critical pump rate.}
	\label{fig:cavity_fluc}
\end{figure}

We now use Eqs.~\eqref{eqs:exp} to calculate cavity field expectations and fluctuations as functions of  the pump rate $\eta\sqrt{N}$ across the NP-SP critical point, as shown in Fig.~\ref{fig:cavity_fluc}(a). In the NP, $|\langle\hat{a}\rangle|$ takes a small finite value because the harmonic trap breaks the $\mathbb{Z}_2$ symmetry.
The NP-SP transition around the critical value $\eta_c\sqrt{N}\approx480$ ($2\pi\times121$kHz) is signaled by a sharp increase in $|\langle\hat{a}\rangle|> 0$, accompanied by enhanced fluctuations $\langle\delta \hat{a}^2\rangle$. 
This is consistent with theoretical \cite{nagy11} as well as experimental~\cite{brennecke13} results. 

In MCTDH-X, these large fluctuations in the vicinity of the critical point also manifest in the convergence of the results with respect to the number of orbitals [cf. Eq.~\eqref{eq:occupancy}]. Fig.~\ref{fig:cavity_fluc}(b) shows that in the critical region, at least $M=5$ orbitals are occupied and required to describe the system.
The inadequacy of the orbital number manifests itself as kinks in the cavity fluctuation curve around $\eta\sqrt{N}=390$ ($2\pi\times98.3$kHz) and 570 ($2\pi\times144$kHz).
Additionally, three orbitals have a sizable occupation in the SP, corresponding to the self-organization pattern of the atoms. Similar results are seen for higher particle numbers.
To summarize, the LRI-MCTDH method is able to  describe the NP and the NP-SP transition in a quantitative manner for a realistic experimental system.

\subsection{Mapping to the Bose-Hubbard model}\label{subsec:BH}
\label{sec:BH_model}

In the self-organized SP, more interesting features can be observed, for example, a lattice switching between the two configurations of the broken $\mathbb{Z}_2$ symmetry [cf. Figs.~\ref{densities}(c I) and (d I)] and the vanishing of superfluidity [cf. Figs.~\ref{densities} (e I)-(e IV)]. These phenomena reflect emergent correlations in the steady state of the system and can be better understood by a mapping of the system to an effective Bose-Hubbard (BH) model.

In the SP, the atomic density distribution has peaks at the lattice sites $x_i=i\pi/k_c$, with $i$ even (odd) integers for the even (odd) lattice configuration. Assuming that only one Wannier function $W_i(x) \equiv W(x-x_i)$ is required for each lattice site, 
the atomic field operator $\hat{\Psi}(x)$ can be rewritten 
as $ \hat{\Psi}(x)=\sum_{i}W^*_{i}(x)\hat{c}_{i}$, where the $\hat{c}_i$ are the annihilation operators for bosons in the Wannier functions $W^*_i(x)$.
With this substitution, the cavity-BEC Hamiltonian [Eq.~\eqref{hamil}] can be approximated by an effective Bose-Hubbard (BH) model with fixed total number of particles:
\begin{equation} \label{BH_Ham}
\hat{\mathcal{H}}_{\textrm{BH}} = -t\sum_{\langle i,j\rangle} \big( \hat{c}^\dagger_i\hat{c}_j+\text{h.c.} \big) + \frac{U_s}{2}\sum_i \hat{n}_i^2+\sum_i \mu_i \hat{n}_i,
\end{equation}
where the first summation only runs over neighboring lattice sites, $\hat{n}_i = \hat{c}^\dagger_i\hat{c}_i$ gives the number of particles at site $i$, and constant terms are omitted.
Expressed in dimensionful units, the hopping strength $t=\frac{\hbar^2}{2m}\int \mathrm{d}x W(x)\partial_x^2 W(x+\frac{2\pi}{k_c})$ originates from the kinetic term, 
$U_s = g \int \mathrm{d}x |W(x)|^4 $ is the on-site interaction, and $\mu_i = \int \mathrm{d}x |W(x-x_i)|^2 V_\text{trap}(x)$ plays the role of a chemical potential.

The  Wannier function $W(x)$ is determined by the cavity-induced interaction. To obtain it, we treat the long-range interaction in Eq.~\eqref{eq:LRI_appr} within a Hartree approximation:
 $\hat{\Theta}^2 = 2\langle\hat{\Theta}\rangle\hat{\Theta}-\langle\hat{\Theta}\rangle^2$, where the operator
$\hat{\Theta}$ is defined in Eq.~\eqref{def_theta}. 
The cavity-induced two-body interaction is now approximated by a one-body potential.
When this potential is deep, the atoms are strongly confined and effectively experience local harmonic potentials of the form
$\frac{1}{2} m\Omega^2(x-x_i)^2$, with $\Omega=\eta k_c\sqrt{2N\hbar/m|\Delta_c|}$. We thus approximate the Wannier function by the ground state of the harmonic potential $W(x) \approx (m\Omega/\pi\hbar)^{1/4}\exp(-m\Omega x^2/2\hbar)$.
Under this approximation, the hopping strength, on-site interaction, and local chemical potential in the Bose-Hubbard model [Eq.~\eqref{BH_Ham}] are given by
\begin{subequations}
\begin{align}
t\approx&\frac{\hbar^2}{2m}\exp\left(-\frac{m\Omega}{\hbar} \frac{\pi^2}{k_c^2}\right)\left[-\frac{m\Omega}{2\hbar} + 
\left(\frac{m\Omega\pi}{\hbar k_c}\right)^2\right]\label{BH_para_t}\\
U_s \approx& g \sqrt{\frac{m\Omega}{2\pi\hbar}} \label{BH_para_us}\\
\mu_i \approx& \frac{m}{2}\frac{\omega_x^2\pi^2}{k_c^2}i^2 \equiv \frac{1}{2}\omega^2i^2.
\end{align}
\end{subequations}
These results are similar to the ones obtained in Ref.~\cite{bakhtiari15} for the two-dimensional case. This similarity stems from the fact that the interaction between the atoms is infinite-range, which renders the dimensionality of the problem unimportant.

\subsection{Self-organized superfluid - self-organized Mott insulator transition}\label{subsec:SSF-SMI}

\begin{figure}[t]
	\centering
		\includegraphics[width=\columnwidth]{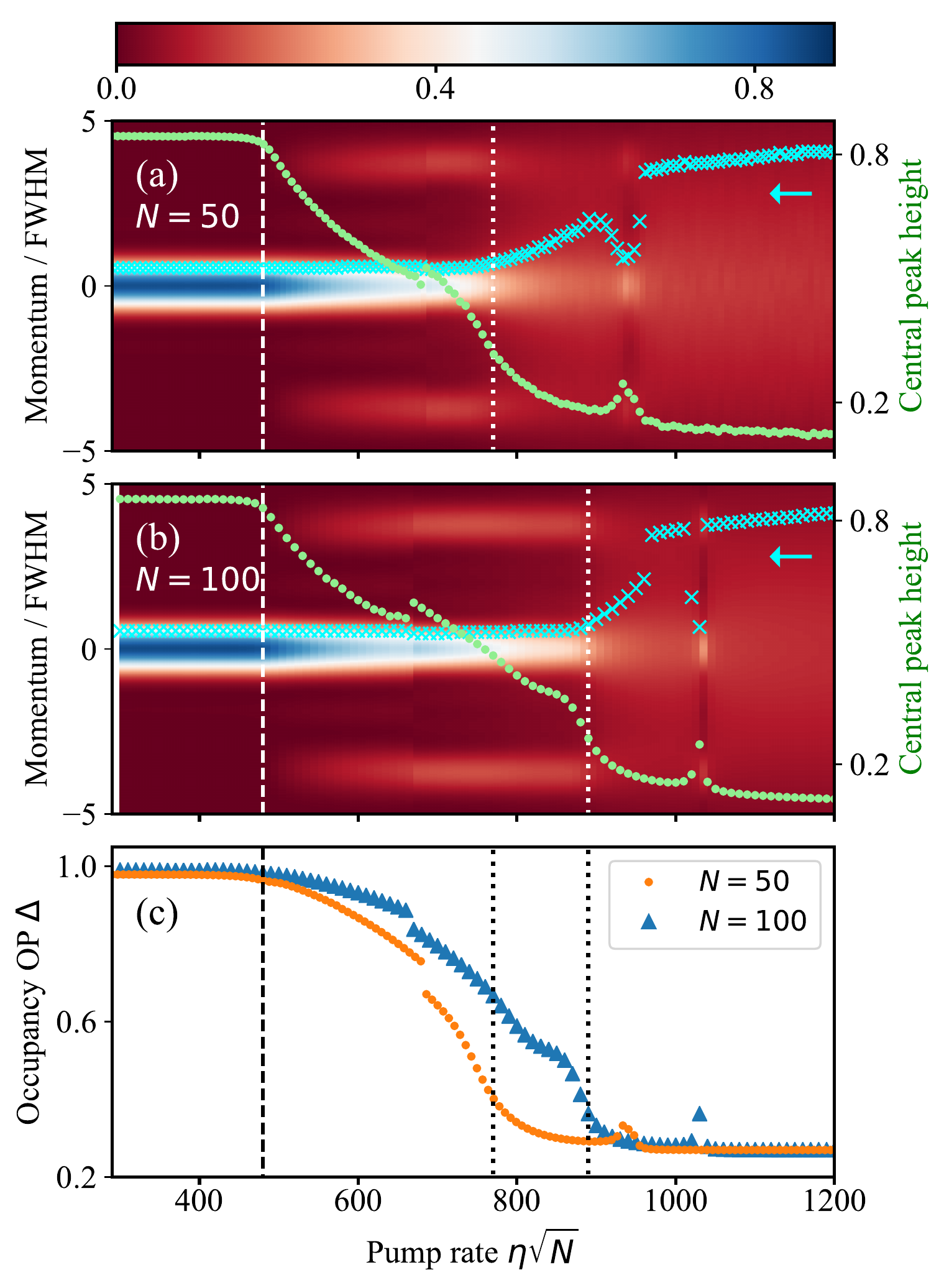}
	\caption{Density distribution in momentum space $\tilde{\rho}(k)$ of (a) $N=50$ and (b) $N=100$ atoms in the lowest energy steady state of the coupled cavity-BEC system as a function of the pump rate $\eta\sqrt{N}$. The height $\tilde{\rho}(k=0)$ and the full width at half maximum (FWHM) of the central peak are superimposed as light green dots and light blue crosses, with axes on the right and left, respectively. Both transitions, as well as the reentrance of superfluidity in the SMI phase, can be clearly seen. 
	(c) The occupancy order parameter $\Delta$ [cf. Eq~\eqref{eq:occ_op}] as a function of pump rate $\eta\sqrt{N}$ simulated with (orange dots) $N=50$ and (blue triangles) $N=100$ atoms. It has similar behaviors as the central peak height in momentum space. The vertical white dashed (dotted) lines in (a) and (b) indicate the NP-SP (SSF-SMI) phase boundary, and they are repeated as black lines in (c). }
	\label{kexpansion}
\end{figure}
A transition from a self-organized superfluid phase (SSF) and a self-organized Mott insulator phase (SMI) in the SP has already been observed in experiments~\cite{klinder15}.
In MCTDH-X simulations, this transition clearly manifests itself in both the momentum space density distribution $\tilde{\rho}(k)$ [see Fig.~\ref{kexpansion}(a)] and in the orbital occupancies [see Figs.~\ref{kexpansion}(c) and \ref{occupancy}]. 

Typically, measures of superfluidity can be extracted from the momentum space distribution, either from the height $\tilde{\rho}(k=0)$~\cite{kato08, wessel04} or the full width at half maximum (FWHM)~\cite{klinder15, greiner02, wessel04} of the central peak.
The transition to the Mott phase is described in Figs.~\ref{densities}(d) and (e).
As the pump rate increases, the system gradually loses its superfluid features both in the momentum space distribution and in the correlation functions. In momentum space [Fig.~\ref{densities}(d II)], the two peaks located at $k=\pm k_c$ slowly merge with the central peak. The off-diagonal terms of $g^{(1)}$  [Fig.~\ref{densities}(d III)] and the diagonal terms of $g^{(2)}$  [Fig.~\ref{densities}(d IV)] both show that the system is now far from the Gross-Pitaevskii mean-field limit. 
Deep in the SMI phase, the momentum distribution is almost flat [Fig.~\ref{densities}(e II)]. In $g^{(1)}$, the diagonal terms are unity, while the off-diagonal terms are zero [Fig.~\ref{densities}(e III)]; in $g^{(2)}$, the diagonal terms are antibunched, while the off-diagonal terms are unity [Fig.~\ref{densities}(e IV)].
All these features indicate that the non-local correlation between the peaks in position space has completely vanished and the hopping term in the BH model is $t=0$.
In the position space density distribution[Figs.~\ref{densities}(d I) and (e I)], the peaks become more and more separated from each other.

An alternative measure is the occupancy order parameter $\Delta$, which is defined through the orbital occupancies $\rho_i$,
\begin{eqnarray}\label{eq:occ_op}
\Delta=\sum_{i=1}^M\rho_i^2,
\end{eqnarray}
and can be detected in experiments by single-shot measurements~\cite{chatterjee17}. As shown in Fig.~\ref{kexpansion}(c), it has similar behaviors as the momentum space central peak height $\tilde{\rho}(k=0)$ in Figs.~\ref{kexpansion}(a) and (b), and is also able to differentiate between the NP, the SSF and the SMI.

Critical behavior is not observed in our simulations across the  SSF- SMI transition. This is in agreement with existing numerical and experimental results on Bose-Hubbard  systems ~\cite{greiner02, wessel04, kato08, klinder15, bakhtiari15}.  The momentum space density distributions in SP, especially the heights and FWHMs of the central peak, are very similar to those encountered in the pure BH model without optical cavity~\cite{wessel04}. This is a  direct consequence of vanishing cavity fluctuations deep in SP, as shown previously in Fig.~\ref{fig:cavity_fluc}. Due to the lack of critical behavior, the boundary between SSF and SMI is ambiguous. One criterion is to choose the boundary as the point where the FWHM of the central peak starts to increase~\cite{wessel04}, roughly at $\eta_\text{mi}\sqrt{N}=770$ ($2\pi\times194$kHz) in our simulations. During the transition from SSF to SMI, two unexpected behaviors can be seen in the density distribution $\tilde{\rho}(k)$ [see Fig.~\ref{kexpansion}(a)]. The switching between the even and odd lattice configurations results in the singularity at around $\eta\sqrt{N}=680$ ($2\pi\times171$kHz), and a reentrance of superfluidity occurs at around $\eta\sqrt{N}=930$ ($2\pi\times234$kHz). Both of them are direct results of the presence of the harmonic trap and will be explained in detail in Sections ~\ref{subsec:SMI} and \ref{subsec:reentrance}, respectively.

The subdivision of SSF and SMI is not only detected in the MCTDH-X simulation, but is also an emergent feature of the BH toy model [Eq.~\eqref{BH_Ham}]~\cite{fisher89}. 
The BH model predicts a transition from a superfluid phase to a Mott insulator phase at $t/U_s\approx0.15/dq$, where $d=1$ is the dimension of the system and $q$ is the filling factor~\cite{teichmann09}. For our system with $N=50$ particles, the density profile implies a filling factor of $q\approx15$. Inserting the parameters of our system, this transition is predicted to occur at approximately $\eta_\text{bh}\sqrt{N}\approx 550$ ($2\pi\times139$kHz), which is approximately 30\% smaller than $\eta_\text{mi}$. This discrepancy can be attributed to the fact that the Gaussian orbitals are not a perfect approximation of the Wannier functions [see Appendix \ref{sec:accuracy_BH}].

The transition between SSF and SMI is intrinsically beyond Gross-Pitaevskii mean-field, so its critical pump rate $\eta_\text{mi}$ is expected to change when the atom number and parameters change [see Eq.~\eqref{numberscale}]. This change is captured by the BH model.
As the atom number increases, the filling factor $q$ increases accordingly. This renders a decrease in the critical value of $t/U_s$. Combining this observation with the expression for $t$ and $U_s$ from the BH model [Eqs.~\eqref{BH_para_t} and \eqref{BH_para_us}], we find that the critical pump rates at two different particle numbers $N$ and $N'$ are related by
\begin{eqnarray}\label{BH_critical_increase}
\eta'_\text{mi}\sqrt{N'}-\eta_\text{mi}\sqrt{N}\approx \frac{k_c}{\pi^2}\sqrt{2|\Delta_c|}\ln\left(\frac{N'}{N}\right).
\end{eqnarray}
It is clear that the critical pump rate will increase as the atom number increases.

To obtain the dependence of $N$ of $\eta_\text{mi}$ in MCTDH-X, we simulate a system of $N'=100$ atoms and compare the results with those from $N=50$ atoms.
Though no changes are seen in the position space distribution $\rho(x)$,  significant changes are observed in the momentum space distribution $\tilde{\rho}(k)$, which is shown in Fig.~\ref{kexpansion}(b). Two satellite peaks emerge at $k=\pm k_c$ and the system enters the SP/SSF at the same pump rate value $\eta_c\sqrt{N}$ as with $N=50$ atoms, underscoring the mean-field nature of the NP-SP transition. In contrast, the  SSF-SMI transition occurs at a larger value of the pump rate $\eta'_\text{mi}\sqrt{N'}=890$ ($2\pi\times224$kHz) as opposed to the pump rate $\eta_\text{mi}\sqrt{N}=770$  ($2\pi\times194$kHz) for $N=50$. This increase is qualitatively consistent with the one predicted by the BH model Eq.~\eqref{BH_critical_increase}.  Since the critical pump rate increases logarithmically, the critical pump rate should remain at the same order of magnitude $\eta_\text{im}\sqrt{N}\approx 2000$ ($2\pi\times500$kHz) even in a usual experiment, where the number of atom in the original two-dimensional system is effectively roughly $N\sim10^3$~\cite{landig16}.
We conclude that just like in the Bose-Hubbard model, the cavity-BEC system can host a Mott phase for any particle number for appropriate cavity detuning and pump rate.

\subsection{ Lattice switching }\label{subsec:SMI}

\begin{figure}[t]
	\centering
		\includegraphics[width=\columnwidth]{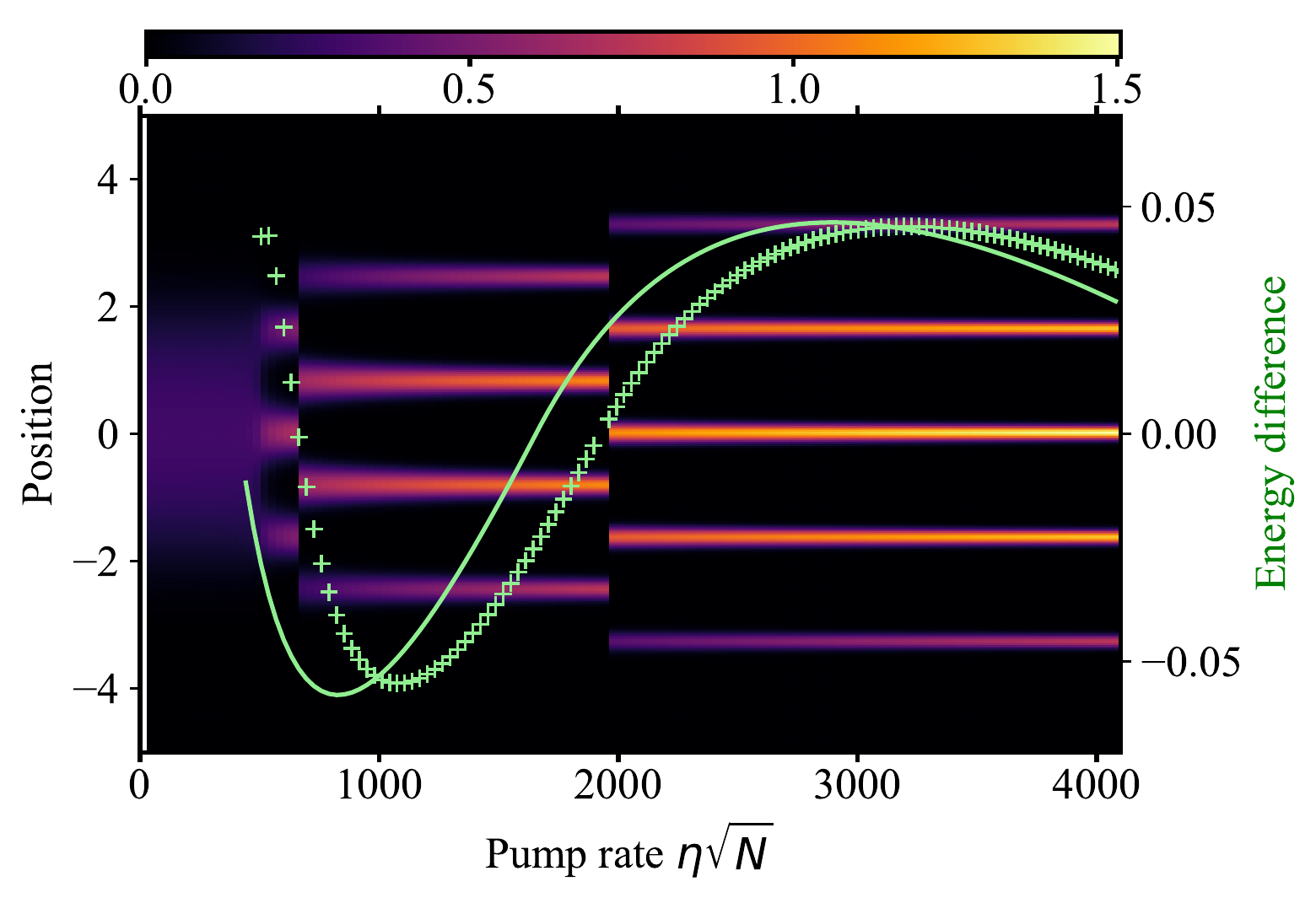}
	\caption{Density distribution $\rho(x)$ of the atoms in the lowest energy steady state for the coupled cavity-BEC system as a function of the pump rate $\eta\sqrt{N}$. The transition from the NP to the SP is clearly seen while the transition from the SSF to the SMI is not visible. However, an expansion of the atomic cloud and the accompanying lattice switching between the two configurations of the broken $\mathbb{Z}_2$ symmetry can be seen.
	The energy difference between these two lattice configurations, $\Delta E_n/N$, is superimposed, with the numerical results from MCTDH-X as green crosses and the analytical results from Bose-Hubbard model [Eq.~\eqref{BH_energydiff}] as a green line. A positive (negative) $\Delta E_n$ value represents that the even lattice has lower (higher) energy than the odd lattice. As expected, the lattice configuration switches every time $\Delta E_n$ vanishes. The results are invariant under a change of atom number $N$.
	}
	\label{xexpansion}
\end{figure}

\begin{figure}[t]
	\centering
	\includegraphics[width=\columnwidth]{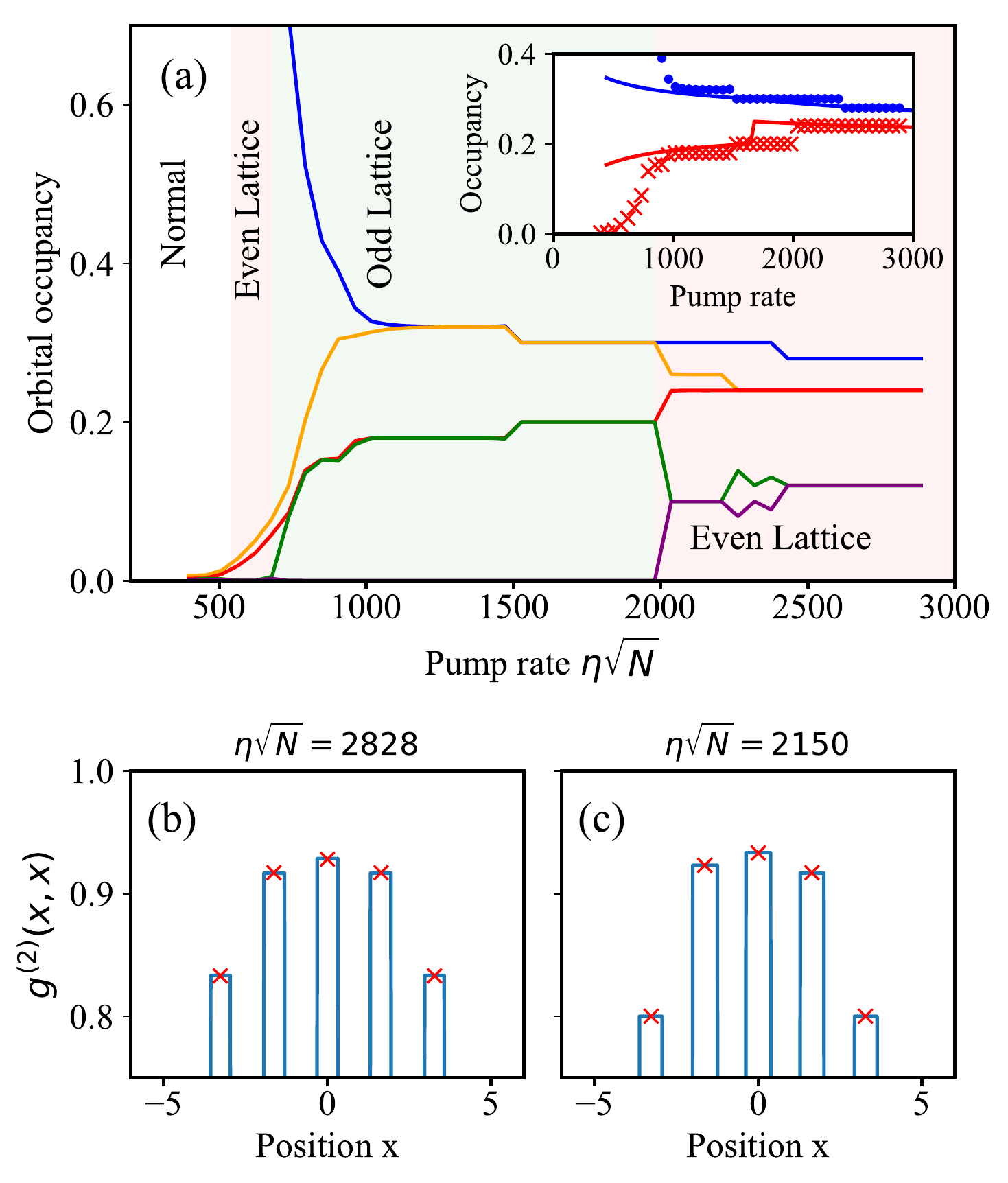}
	\caption{ (a) Occupancies of the $M=5$ orbitals for $N=50$ particles as functions of the pump rate $\eta\sqrt{N}$. The background color shows whether the ground state has an even (pink) or odd (light green) lattice configuration. Inset: Occupancies of the first and third orbitals ($\rho_1$ and $\rho_3$) obtained from MCTDH-X simulations (markers) compared with the proportions of atoms in the corresponding peaks calculated from the analytical BH model in Eq.~\ref{eq:atom_num_site} (solid lines). For the even lattices, these two orbitals correspond respectively to the peaks at $x=0$ ($n_0/N$ in BH model) and $x=2\pi/k_c$ ($n_2/N$). For the odd lattices, they correspond to the peaks at $x=\pi/k_c$ ($n_1/N$) and $x=3\pi/k_c$ ($n_3/N$). (b) and (c) The diagonal of the two-body correlation function $g^{(2)}(x,x)$ in two self-organized Mott insulator states. The states are simulated at (b) $\eta\sqrt{N}=$ 2828 ($2\pi\times713$kHz) and (c) 2150 ($2\pi\times542$kHz) with $N=50$. The occupancies of the five orbitals in (b) the first state are 28\%, 24\%, 24\%, 12\%, 12\%, while in (c) the second state they are 30\%, 26\%, 24\%, 10\%, 10\%. The blue lines are simulation results and the red crosses are analytical results obtained from Eq.~\eqref{eq:2cf}.}
	\label{occupancy}
\end{figure}
We now focus on the spatial profile of the condensate as a function of the pump rate (Fig.~\ref{xexpansion}). Other than the expected transition from NP to SP at $\eta_c\sqrt{N}=480$ ($2\pi\times121$kHz) and the narrowing of the peaks as the pump rate increases, we can further see a switching between the even and odd lattice configurations. Using the BH model, we now show that this switching stems from the presence of the harmonic trap, which explicitly breaks the $\mathbb{Z}_2$ symmetry and lifts the degeneracy of the two configurations. 
We also note that the scenario where the harmonic trap leads to different domains in different positions~\cite{batrouni02} is not observed in our system, because the number of occupied lattice sites is too small near the boundary between SSF and SMI.

In the atomic limit of the BH model ($t=0$), we can calculate the lowest energy configurations as a function of the dimensionless parameter $\xi \equiv NU_s/\omega^2\propto\sqrt{\eta}$. The detailed calculation is shown in Appendix~\ref{sec:BH_peaks}, where the configuration and the energy of the lowest energy steady state are obtained. The BH model predicts the dependence on $\xi$ of the number of occupied lattice sites $n_\text{site}$. In the range of pump rates relevant for our simulations, $n_\text{site}$ is given by
\begin{eqnarray}
n_\text{site}=
\begin{cases}
3, \quad 8<\xi\le20\quad&(66<\eta\sqrt{N}\le415) \\
4, \quad 20<\xi\le40\quad&(415<\eta\sqrt{N}\le1659) \\
5, \quad 40<\xi\le70\quad&(1659<\eta\sqrt{N}\le5080).
\end{cases}
\end{eqnarray}
As the pump rate $\eta\sqrt{N}$ and therefore $\xi$ increase gradually, the number of occupied lattice sites $n_\text{site}$ increases by steps of one at a time. Each step is accompanied by a switching between the even and odd lattices. This is consistent with the simulation results in Fig.~\ref{xexpansion} except for a shift in the pump rate, which we attribute to the fact that the Gaussian orbital is not a perfect approximation for the Wannier orbital [see Appendix~\ref{sec:accuracy_BH}].

The BH model can also predict the energies of higher energy states. For example, it can be used to calculate the energy differences $\Delta E_{n_\text{site}}$ between the two lowest energy steady states at different pump rates [see Appendix~\ref{sec:BH_peaks}],
\begin{eqnarray}\label{BH_energydiff}
\frac{\Delta E_n}{N\omega^2} =-\frac{(n^3-n-3\xi)[n(n+1)(n+2)-3\xi]}{18n(n+1)\xi}.
\end{eqnarray}
For the sake of clarity, the notation for the number of occupied sites $n_\text{site}(\xi)$ is simplified as $n$ in this formula. This energy difference is also accessible in MCTDH-X. The analytical and numerical results for $\Delta E_n$ are compared in Fig.~\ref{xexpansion}, and they agree with each other except for a shift [see Appendix~\ref{sec:accuracy_BH}].
	
With MCTDH-X and the BH model, we can probe the number of atoms localized at each lattice sites both numerically and analytically.
As the atom peaks are locally correlated deep in the Mott phase, for any given position $x=x_0$, there should be at most one macroscopically-occupied orbital which is localized nearby.
The diagonal term of the two-body correlation function $g^{(2)}(x,x)$ [see Eq.~\eqref{eq:def_g2}] establishes a direct connection between the density peaks and the occupancy $\rho_i$ of this orbital [see Appendix~\ref{sec:g2} and cf. Eq.~\eqref{eq:occupancy}],
\begin{eqnarray}\label{eq:2cf}
g^{(2)}(x_0,x_0) =1-\frac{1}{N\rho_i}.
\end{eqnarray}
The orbital occupancy $\rho_i$, on the other hand, reflects the fraction of the total number of atoms in the corresponding peaks located at $x=x_0$.
This can be substantiated with a comparison of the orbital occupancies $\rho_i$ from simulations with the atom fractions in the corresponding peaks using the BH model
[see Appendix~\ref{sec:BH_peaks}],
\begin{eqnarray}\label{eq:atom_num_site}
\frac{n_i}{N}= \frac{1}{n_\text{site}}+\frac{n_\text{site}^2-3i^2-1}{6\xi}.
\end{eqnarray}
The orbital occupancies of the five orbitals as functions of pump rate are shown in Fig.~\ref{occupancy}(a), and they are compared to the analytical results Eq.~\eqref{eq:atom_num_site} in the inset. 
The analytical and numerical results agree with each other except for a horizontal shift [see Appendix~\ref{sec:accuracy_BH}], confirming our claim.

Combining the results from Eq.~\eqref{eq:2cf} and our claim, the value of $g^{(2)}(x,x)$ at each peak is directly related to the number of atoms $N_i=N\rho_i$ therein.
The simulation results and the analytical results from Eq.~\eqref{eq:2cf} of two SMI states simulated at $\eta\sqrt{N}=2828$ ($2\pi\times713$kHz) and 2150 ($2\pi\times542$kHz) are compared in Figs.~\ref{occupancy}(b) and (c), respectively. The calculations faithfully reproduce the MCTDH-X results. In particular, the asymmetry encountered in the $ x=\pm 2\pi/k_c$ peaks in the state simulated at $\eta\sqrt{N}=2150$ ($2\pi\times542$kHz) is a consequence of the indivisibility of the atoms and is also visible in the second and third orbital occupancies in Fig.~\ref{occupancy}(a).

\subsection{Reentrance of superfluid features on the Mott insulator side of the SSF-SMI boundary}\label{subsec:reentrance}
Besides the switching between the even and odd lattices, the harmonic trap also induce a reentrance of correlation and superfluidity in the SMI phase. This can be seen around the pump rate $\eta\sqrt{N}=930$ ($2\pi\times234$kHz) in the momentum space density distributions in Fig.~\ref{kexpansion}  for $N=50$ atoms.
As a global effect, it is not captured by a local density approximation~\cite{bergkvist04,laura06}.
The reentrance can also be explained by the physics of the BH model.
Deep in the SMI, as $\eta$  increases,  Eq.~\eqref{BH_para_us} shows that the on-site repulsive interaction between atoms $U_s$ becomes progressively more significant than local chemical potentials $\mu_i$ induced by the harmonic trap.
Consequently, the atoms minimize their energy by populating lattice sites further from the minimum of the trap instead of congregating at its center. Due to the finite number of atoms and their indivisibility, only one atom from each of the inner peaks hops outwards at a time.
This process also occurs in the SMI region close  to the SSF-SMI phase boundary -- the fact that the hopping strength $t$ is still comparable to the on-site interaction $0.01U_s$ [cf. Section~\ref{subsec:SSF-SMI}] leads to the reentrance of correlations and superfluidity.

As the pump rate $\eta$ increases across a certain critical pump rate $\eta^*$, the system gradually transitions from a state $|1\rangle$, where more atoms sit in the inner lattice sites to another state $|2\rangle$, where they move outwards. The energy difference $2\epsilon$ between these two states is a function of $U_s$ and $\omega$, and, hence, of $\eta$. The correlation strength among the lattice sites is determined by the ratio between the energy difference $\epsilon(\eta)$ and the hopping strength $t(\eta)$ [see Appendix~\ref{sec:append_re}]. Far away from $\eta^*$, the energy difference is much larger than the hopping strength $|\epsilon|\gg t$, so the steady state of the system is dominated by either $|1\rangle$ or $|2\rangle$. However, in the vicinity of $\eta^*$, the energy difference is comparable  to or smaller than the hopping strength $|\epsilon|<10t$, and the steady state is a superposition of many states, including $|1\rangle$ and $|2\rangle$. This superposition reestablishes coherence between the lattice sites and causes the reentrance of superfluidity in the SMI. An example with $n_\text{sites}=4$ lattice sites is calculated and discussed in detail in Appendix~\ref{sec:append_re}. 

\begin{figure}[t]
	\centering
	\includegraphics[width=\columnwidth]{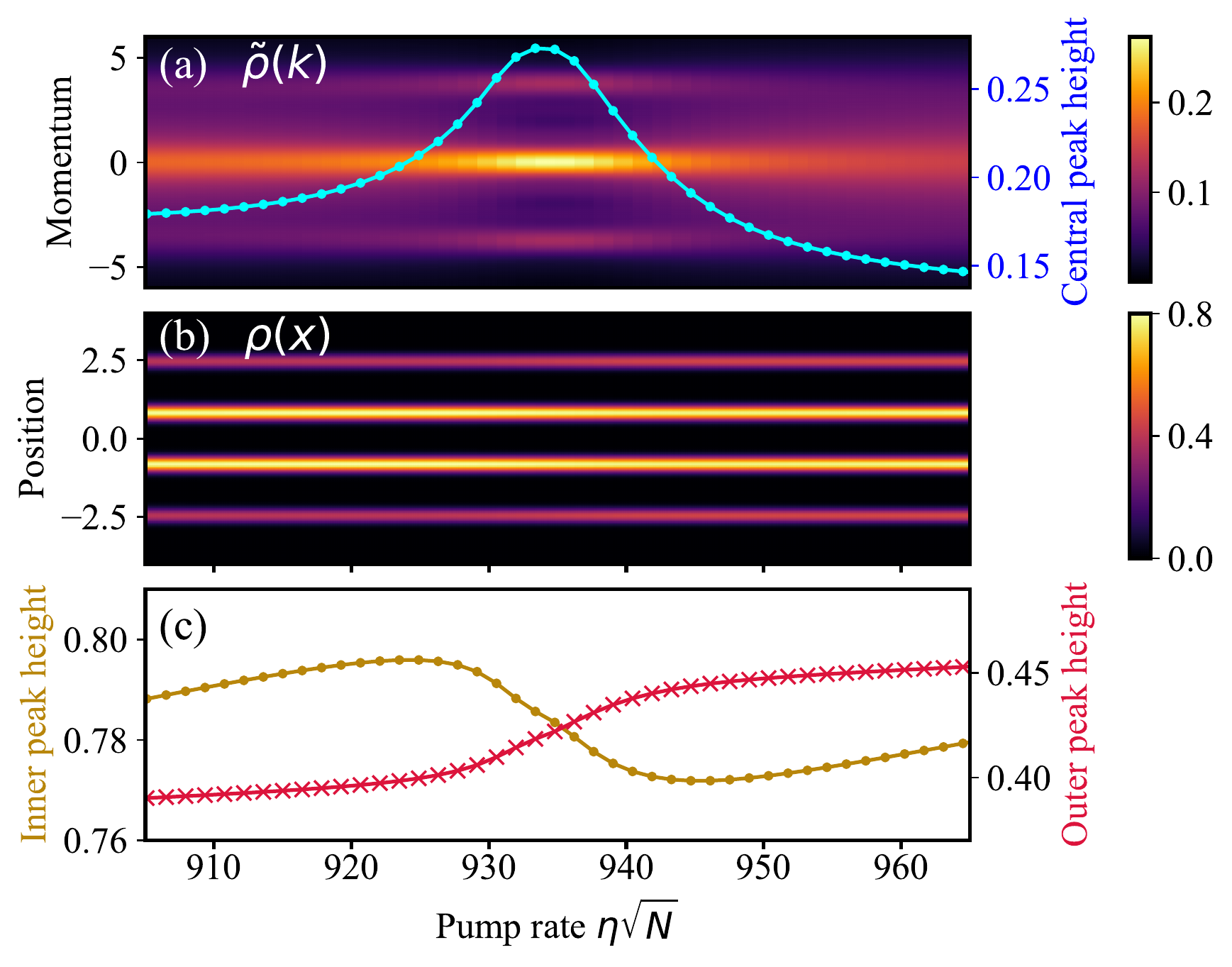}
	\caption{Density distributions in position and momentum spaces as functions of the pump rate $\eta\sqrt{N}$ in the reentrance region for $N=50$ atoms. (a) Density distributions in momentum space $\tilde{\rho}(k)$. Superimposed is the height of the central peak $\tilde{\rho}(k=0)$. (b) Density distributions in position space $\rho(x)$. (c) The heights of the inner peaks $\rho(x=\pm\pi/k_c)$ (brown dots) and outer peaks $\rho(x=\pm3\pi/k_c)$ (red crosses) of the position space density distribution. The rapid change in the peak heights imply an atom hops from each of the inner peaks into an outer peak in the reentrance region, resulting in a reconstruction of correlations between peaks.}
	\label{fig:reentrance}
\end{figure}

The MCTDH-X simulation results confirm this mechanism. The density distributions in momentum space and position space around the reentrance region for $N=50$ atoms are shown in Figs.~\ref{fig:reentrance}(a) and (b),
respectively. As the central peak height in momentum space $\tilde{\rho}(k=0)$ enhances, in position space the inner peaks shrink and the outer peaks grow rapidly in height [see Fig.~\ref{fig:reentrance}(c)]. This implies that an atom is moving from each of the inner to outer peaks when the superfluid features reenter, which is consistent with the discussion above.

The reentrance also manifests itself in other observables like the one-body and two-body correlation functions in both position and momentum space.
It is also seen for $N=100$ atoms [see Fig.~\ref{kexpansion}(b)], where it occurs at $\eta\sqrt{N}=1050$ ($2\pi\times265$kHz). 
There is no reentrance at higher pump rates as the hopping $t$ decays exponentially.

%%%%%%%%%%%%%%%%%%%%%%

%%%%%%%%%%%%%%%%%%%%%%%%%%%%%%%%%-DISCUSSION-%%%%%%%%%%%%%%%%%%%%%%%%%%%%%%%%%

\section{Conclusions}\label{sec:conclusion}

In summary, we have studied the steady-state properties of laser-pumped harmonically-trapped weakly-interacting bosons coupled to a red-detuned dissipative high-finesse optical cavity.
We explicitly considered experimentally relevant parameters for our simulations, permitting a direct comparison of our results with state-of-the-art experimental setups.

We showed that the bosonic atoms manifest three different phases as a function of the strength of the pump laser for a fixed cavity detuning: a normal phase (NP) where the atoms form a BEC, a self-organized superfluid (SSF) phase and a self-organized Mott-insulator (SMI) phase.
In the steady state, each phase is characterized by distinct features in its momentum space density distribution and its correlation functions.
In the NP, the atoms form an almost perfect condensate, i.e., only a single orbital is macroscopically occupied. The density distribution in position and momentum space displays a single peak, whose shape depends very weakly on the pump rate.
Around the critical point between the NP and the SSF phase, the cavity fluctuations are significantly enhanced even for a small number of atoms, in accordance with the expected divergence of the fluctuations for a symmetry-breaking quantum phase transition.
The system then becomes fragmented as it enters the SSF phase.
In the SSF phase, a fraction of the atoms form clusters in position space.
Correspondingly, the momentum space density distribution is characterized by two satellite peaks located at the wave vectors $k = \pm k_c$, indicating a strong correlation between the atom clusters.
As the pump rate increases further, the system smoothly transitions from the SSF into the SMI phase without exhibiting critical behavior.
The transition is signaled by changes in the height and width of the central peak in the momentum space as well as in the occupancy order parameter.
This is in agreement with previous numerical and experimental studies of the Mott transition in standard Bose-Hubbard models. 
Deep in the SMI phase, the atomic state is highly fragmented and locally correlated.  

Surprisingly, we find two novel features inside the superradiant phase:
i) an expansion and switching of the self-organized lattice between the two configurations of the broken $\mathbb{Z}_2$ symmetry, 
ii) a previously unseen reentrance of superfluid features in the SMI phase.
The observed phenomena can be qualitatively explained via a mapping of the cavity-BEC system to the Bose-Hubbard model, and they are attributed to the competition between the harmonic trapping potential and the contact interaction energy between the atoms.
Slight discrepancies are attributed to the inaccuracy of our approximation for the Wannier function used to obtain the Bose-Hubbard model.

Our work shows that the combination of numerical MCTDH-X simulations and analytical results from effective Hubbard models offer a comprehensive approach to study correlated phases of matter in cavity-cold atomic gases setups. 
Additional prospects for application of this methodology range from cavity-coupled multicomponent bosons --- where exploratory studies have already been conducted~\cite{axel18} --- to ultracold fermionic gases and other engineered light-matter systems with multimodal cavities or with the presence of additional optical lattices.
The easy incorporation of time-dependence provided by the MCTDH-X algorithm further enables a highly controlled and systematic way to investigate the dynamical behaviors of both quantum quenches and Floquet systems.

%%%%%%%%%%%%%%%%%%%%%%%%%%%%%%%%%-ACKNOWLEDGEMENTS-%%%%%%%%%%%%%%%%%%%%%%%%%%%%%%%%%

\acknowledgments

We acknowledge the financial support from the Swiss National Science Foundation (SNSF), Mr. Giulio Anderheggen, the Austrian Science Foundation (FWF) under grant No. F65 (SFB ‘Complexity in PDEs’), grant No. F41 (SFB ‘ViCoM’), the Wiener Wissenschafts- und TechnologieFonds (WWTF) project No. MA16-066 (‘SEQUEX’). We also acknowledge the hospitality of the Wolfgang-Pauli-Institut and the computation time on the ETH Euler and HLRS Stuttgart HazelHen clusters.

%%%%%%%%%%%%%%%%%%%%%%%%%%%%%%%%%%%%-APPENDIX-%%%%%%%%%%%%%%%%%%%%%%%%%%%%%%%%%%%%%

 \appendix

\section{Numerical method}\label{sec:mctdhx}
In this work, we use the Multiconfigurational Time-Dependent Hartree method for indistinguishable particles~\cite{axel16,alon08,fasshauer16} (encoded in the software MCTDH-X~\cite{ultracold}) to accurately investigate the ground state of the problem. We consider a general Hamiltonian containing at most two-body operators:
\begin{align} 
\hat{\mathcal{H}}&=\int \mathrm{d}x \hat{\Psi}^\dagger(x) \left\{\frac{p^2}{2m}+V(x)\right\}\hat{\Psi}(x) \nonumber\\
&+\frac{1}{2}\int \mathrm{d}x \hat{\Psi}^\dagger(x)\hat{\Psi}^\dagger(x')W(x,x')\hat{\Psi}(x)\hat{\Psi}(x').
\end{align}
Here $V(x)$ represents the one-body potential and $W(x,x')$ the two-body interactions.
The numerical method is based on the following ansatz for the many-body wave function
\begin{eqnarray}
	|\Psi\rangle=\sum_{\mathbf{n}} C_\mathbf{n}(t)\prod^M_{k=1}\left[ \frac{(\hat{b}_k^\dagger(t))^{n_k}}{\sqrt{n_k!}}\right]|0\rangle ,
\end{eqnarray}
where $N$ is the number of atoms, $M$ is the number of single-particle wave functions (orbitals) and $\mathbf{n}=(n_1,n_2,...,n_k)$ gives the 
number of atoms in each orbital, i.e. $\sum_{k=1}^M n_k=N$. The vacuum is denoted by $|0\rangle $ and the time-dependent operator 
$\hat{b}_k^\dagger$ creates one atom in the $i$-th working orbital $\psi_i(x)$
\begin{eqnarray}
	b_i^\dagger(t)&=&\int \psi^*_i(x;t)\hat{\Psi}^\dagger(x;t) \mathrm{d}x \\
	\hat{\Psi}^\dagger(x;t)&=&\sum_{i=1}^M b^\dagger_i(t)\psi_i(x;t). \label{eq:def_psi}
\end{eqnarray}
Using the time-dependent variational principle~\cite{TDVM81} one finds the equations for the time-evolution of the coefficients $C_\mathbf{n}(t)$ 
and the working orbitals $\psi_i(x;t)$.

Because of the existence of quasidegenerate states near the ground state energy, the variational procedure could relax the system to some 
metastable states besides the ground state.
To circumvent this problem, we run imaginary time propagations with several (10 or 20) randomly generated initial conditions as ansatz for the wave function and choose the one with the lowest energy as the final state. 
In this way, we can assure that the ground state is reached and, as a byproduct, we can also access the low-lying excited states. The MCTDH-X has been successfully used to study vortex formation in BECs \cite{weiner17}, superlattice switching in parametrically driven condensates\cite{molignini18},  localization in the spin-boson model \cite{wang08} among others.

In the MCTDH-X approach the choice of the number of orbitals $M$ plays an important role for the convergence of the results~\cite{alon08,axel12,fasshauer16}.
In the following, we provide a guideline on how to choose it for systems described by Eq.~\eqref{hamil}. 
When only $M=1$ orbital is used, the solution is reduced to the Gross-Pitaevskii-type mean-field approximation. 
On the contrary, when $M$ goes to infinity the MCTDH-X becomes numerically exact, but a large $M$ is computationally unrealistic. In some cases, however, quasi-exact solutions can be obtained~\cite{axel12}.
To obtain an optimal $M$, we perform imaginary time propagations with different random initial guesses starting from $M=2$ and increase 
$M$ by one if the occupation of the last orbital is non-vanishing in any one of the tests.
The tests show that for our present computations at most as many orbitals as the number of density peaks (atom clusters) in position space are needed, 
except in the vicinity of the NP-SP boundary. Since there are at most five density peaks in our simulations, we set $M=5$ and consequently 
use $N=50$ atoms due to computational limitations. 
We can claim the numerical exactness of our simulations as the higher order orbitals have vanishing occupancies far from the NP-SP boundary.

 \section{Parameters}\label{sec:parameters}

In this paper, all the parameters are expressed in dimensionless units. The scale of energy $\tilde{E}$, scale of length $\tilde{L}$ and scale of frequency $\tilde{\omega}$ can be chosen arbitrarily as long as they satisfy $\tilde{E}=\frac{\hbar^2}{2m\tilde{L}^2}=\hbar\tilde{\omega}$, where $m$ is the mass of the atoms. 
For $^{87}$Rb atoms with mass $1.44\times10^{−25}$kg, we can choose $\tilde{E}=1.67\times10^{-31}$J, $\tilde{L}=480$nm and $\tilde{\omega}=2\pi\times252$Hz. 
In our simulations, all the parameters and results are given in these units.

Most of the parameters are chosen here corresponding to the experimental setup of Ref.~\cite{baumann10}, where the atoms are usually split into multiple two-dimensional slices with approximately $N=1000\sim 2000$ atoms in each slice~\cite{landig16}, $\omega_x=1.0\tilde{\omega}$, $NU_0=213\tilde{\omega}$, $\Delta_c=-4\times10^4\tilde{\omega}$, $k_c=3.84/\tilde{L}$, $\kappa=5158\tilde{\omega}$ and $\eta\sqrt{N}$ varying from 0 to 4000, while the one-dimensional contact interaction strength $Ng=10.01\tilde{\omega}\tilde{L}$ is related to the three-dimensional scattering length according to Ref.~\cite{olshanii98}. The values of these parameters expressed in dimensionful units are shown in Table~\ref{table:conversion}.

\begin{table}
\begin{tabular}{c|c}
	\hline\hline
	Quantity & Value \\
	\hline\hline
	Harmonic trap $\omega_x$ & $2\pi\times 252$Hz \\
	\hline
	Atomic contact interaction $Ng$ & $5.01\times10^{-18}$eV$\cdot$m \\
	\hline
	$NU_0$ & $2\pi\times53.7$kHz\\
	\hline
	Cavity detuning $\Delta_c$ & $-2\pi\times10.1$MHz\\
	\hline
	Cavity mode wave vector $k_c$ & $5.03\times10^7$ m$^{-1}$\\
	\hline
	Cavity loss rate $\kappa$ & $2\pi\times1.30$MHz\\
	\hline
	Pump rate when $\eta\sqrt{N}=2000$ & $2\pi\times504$kHz\\
	\hline\hline
	Energy scale & $1.04\times10^{-12}$eV\\
	\hline
	Length scale & 480nm\\
	\hline\hline
\end{tabular}
\caption{Values of the system parameters, and the energy and length scales in dimensionful units.}\label{table:conversion}
\end{table}

%%%%%%%%%%%%%%%%%%%%%%

\section{Effective potential approach}\label{sec:OBP}

\begin{figure}[t]
	\centering
	\includegraphics[width=\columnwidth]{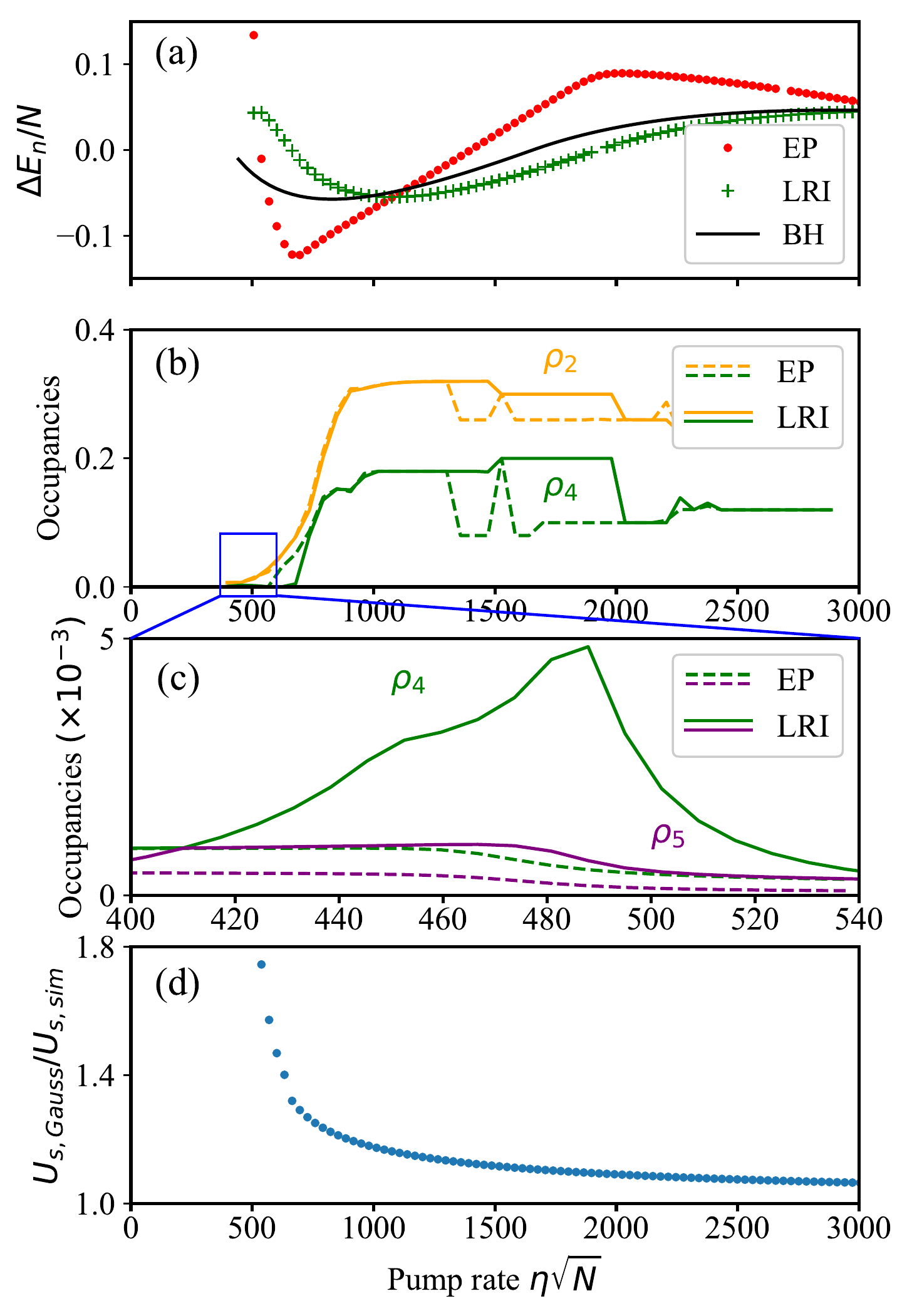}
	\caption{(a) Energy difference between the even and odd lattice configurations $\Delta E_n/N$ as a function of the pump rate $\eta\sqrt{N}$ for an analytical computation using the BH model Eq.~\eqref{BH_energydiff} (solid black line) and the numerical computations (green markers and red dots) using MCTDH-X. The green markers are results from the LRI approach as shown in Fig.~\ref{xexpansion}(a), while the red dots are results from EP approach. 
	(b) and (c) Orbital occupancies as functions of pump rate $\eta\sqrt{N}$ (b) for the whole range of pump rates and (c) in the vicinity of the NP-SP critical point. Only two orbitals are shown in each panel for the sake of clarity. The solid lines show the LRI results as shown in Figs.~\ref{occupancy}(a) and \ref{fig:cavity_fluc}(b), and the dashed lines show the EP results. 
	In SP, the two approaches predict different lattice configurations in certain intervals. Across the NP-SP critical point, a tiny enhancement in the fourth orbital occurs only in the LRI approach.
	(d) Ratio between the values of the on-site interaction $U_s$ calculated by the Gaussian approximation and the simulation results $U_{s,\text{Gauss}}/U_{s,\text{sim}}$. }
	\label{wannier_obp}
\end{figure}

For comparison, we summarize the \emph{effective potential} (EP) approach, where the cavity field is treated in a mean-field fashion. It is assumed to be in a coherent state with no cavity fluctuation and the operator $\hat{a}$ is replaced by a complex number: 
$\hat{a} \mapsto \alpha$~\cite{baumann10,nagy08}. Under this simplification, the evolution of the cavity field is given by the following Heisenberg equation of motion
\begin{eqnarray}
\frac{\partial}{\partial t}\alpha
=i\left[\Delta_c\alpha-U_0\mathcal{B}\alpha-\eta\Theta\right]-\kappa\alpha ,
\end{eqnarray}
where $\Theta=\langle\hat{\Theta}\rangle$ and $\mathcal{B}=\langle\hat{\mathcal{B}}\rangle$ are defined in Eq.~\eqref{eq:b_and_theta},
and the Hamiltonian Eq.~\eqref{hamil} is now dependent on the cavity field $\alpha$ as [cf. Eq.~\eqref{eq:LRI_appr}]
\begin{align}
\hat{\mathcal{H}} & = \int \mathrm{d}x \hat{\Psi}^\dagger(x)\left\{\frac{p^2}{2m}+\frac{g}{2}\hat{\Psi}^\dagger(x)\hat{\Psi}(x)+V_\text{trap}(x)\right\}\hat{\Psi}(x) 
\nonumber \\
& \quad +\hbar U_0\int \mathrm{d}x \hat{\Psi}^\dagger(x) \cos^2(k_c x)|\alpha|^2\hat{\Psi}(x) \nonumber \\
& \quad + \hbar\eta\int \mathrm{d}x \hat{\Psi}^\dagger(x) \cos(k_c x)(\alpha^*+\alpha)\hat{\Psi}(x).
\end{align}
The cavity effectively provides two sinusoidal modifications to the one-body potential of the atoms~\cite{baumann10,maschler08}. In the limit $|\Delta_c|\gg NU_0$, the first contribution is negligible. The second modification favors either the even $i$ or odd $i$ lattice sites depending on the sign of $\Re(\alpha)$. Thus these modifications capture a periodic atomic density distribution with period $2\pi/k_c$, like the LRI approach. 

The results obtained from the LRI and the EP approaches are compared in Figs.~\ref{wannier_obp}(a)-(c). 
In Fig.~\ref{wannier_obp}(a), we compare the energy difference between the two configurations of the broken $\mathbb{Z}_2$ symmetry [cf. Fig.~\ref{xexpansion}]. 
Both simulation results from LRI and EP approaches are compared to the analytical results obtained from the BH model. 
The latter approach also captures the main features like the switching between lattices and the expansion of the atomic cloud, but the energy difference and switching pump rate deviates much more from the BH model than the ones obtained from the former approach.
In Figs.~\ref{wannier_obp}(b) and (c), we compare the orbital occupancies obtained from the LRI and the EP approaches [cf. Figs.~\ref{occupancy}(a) and \ref{fig:cavity_fluc}(b)]. Both approaches give exactly the same occupancies except for the intervals $550<\eta\sqrt{N}<750$ and $1300<\eta\sqrt{N}<2400$, and the vicinity of $\eta_c\sqrt{N}=480$.
In the first case, the difference comes from the fact that the two approaches predict different lattice configurations. In the second case, compared to the LRI approach, the EP approach completely ignores the cavity fluctuation at the critical point $\eta_c$, so the tiny enhancement in the fourth orbital does not occur.

In conclusion, compared to the LRI approach, the EP one completely ignores the cavity fluctuations, but it is able to reproduce the same atomic correlations in the system except in the vicinity of the critical point between NP and SP. 
It estimates different energies for a given state and, consequently, sometimes reveals a state with the opposite lattice symmetry as the lowest energy steady state.

\section{Configurations and energies in the atomic limit}
\subsection{Lattice configurations predicted by the Bose-Hubbard model}\label{sec:BH_peaks}
In the atomic limit, the hopping term in the Bose-Hubbard model [Eq.~\eqref{BH_Ham}] vanishes, so the energies of a state can be written as
\begin{eqnarray}
E_\text{BH} &=& \frac{U_s}{2}\sum_i n_i^2+\frac{\omega^2}{2}\sum_i i^2 n_i \\
&=& \frac{N\omega^2}{2} \sum_i\left(\xi \frac{n_i^2}{N^2}+i^2\frac{n_i}{N} \right),
\end{eqnarray}
where $n_i$ with $i=0,\pm1, \pm2...$ denotes the number of atoms inside each peak, and $n_i=n_{-i}$ due to the symmetry of the harmonic trap. The parameter $\xi=NU_s/\omega^2$ is proportional to the square root of the pump rate $\eta$. To look for the lowest energy steady state, the energy is minimized under the constraint
\begin{equation}
\sum_i n_i = N.
\end{equation}
For simplicity, we consider an infinite number of atoms $N$, such that it can be divided arbitrarily.

The strategy to look for the lowest energy steady state consists of two steps. First, we fix the number of occupied sites $n_\text{site}$ and look for the state with the lowest energy. Second, we compare the results from different $n_\text{site}$ and choose the minimal energy state.

If $n_\text{site}$ is even (odd), the lattice is in the odd (even) lattice configuration and only $n_{\pm1}$, $n_{\pm3}$, ... $n_{\pm (n_\text{site}-1)}$ [$n_0$, $n_{\pm2}$, ... $n_{\pm (n_\text{site}-1)}$] are non-zero.
Using the method of Lagrange multiplier, we can find the number of atoms $n_i$ at each site $i$ as a function of $\xi$ and $n_\text{site}$,
\begin{eqnarray}
\frac{n_{\pm i}}{N} = \frac{1}{n_\text{site}}+\frac{n_\text{site}^2-3i^2-1}{6\xi}.
\end{eqnarray}
This configuration with $n_\text{site}$ lattice sites has an energy of 
\begin{eqnarray}
\frac{E_{n_\text{site}}}{N\omega^2} &=&  \frac{\xi}{2n_\text{site}}+\frac{1}{6}(n_\text{site}^2-1)\nonumber\\
&&-\frac{1}{90\xi}n_\text{site}(n_\text{site}^2-1)(n_\text{site}^2-4),
\end{eqnarray}
and exists only when $\xi$ exceeds a certain value such that $n_{\pm (n_\text{site}-1)}>0$,
\begin{eqnarray}
\xi>\xi_{n_\text{site}}\equiv\frac{1}{3}n_\text{site}(n_\text{site}-1)(n_\text{site}-2).
\end{eqnarray}

Now that we have already obtained the configuration and energy of the lowest energy steady state with a given fixed lattice site number $n_\text{site}$, we can compare the energies $E_{n_\text{site}}$ of different lattice site numbers $n_\text{site}$. In the interval
\begin{eqnarray}
\xi_{n_\text{site}+1}<\xi<\xi_{n_\text{site}+2},
\end{eqnarray}
the steady state with the lowest energy occupies $n_\text{site}$ lattice sites, while the one with the second lowest energy occupies $n_\text{site}+1$ lattice sites. The energy difference between these two steady states $\Delta E_{n_\text{site}} = E_{n_\text{site}+1}- E_{\text{site}}$ is
\begin{eqnarray}
\frac{\Delta E_{n}}{N\omega^2} =-\frac{(n^3-n-3\xi)[n(n+1)(n+2)-3\xi]}{18n(n+1)\xi},
\end{eqnarray}
where the notation $n_\text{site}$ has been simplified as $n$ for the sake of clarity.

\subsection{Two-body correlation function}\label{sec:g2}
In the limit of the BH model, only on-site correlations exist. In MCTDH-X, this local correlation manifests itself in the fact that the orbitals are spatially well-separated from each other. In other words, at any given position $x=x_0$, at most one natural orbital $\phi_i$ is non-zero. 

Combining the definition of the reduced density matrices Eq.~\eqref{eq:def_rho} and the expression of the operator $\hat{\Psi}^\dagger(x)$ in MCTDH-X Eq.~\eqref{eq:def_psi}, we arrive at
\begin{eqnarray}
\rho^{(1)}(x_0,x_0) &=& |\phi_i(x_1)|^2\langle  \hat{b}^\dagger_i\hat{b}_i \rangle=|\phi_i(x_1)|^2N\rho_i\\
\rho^{(2)}(x_0,x_0) &=& |\phi_i(x_1)|^4\langle  \hat{b}^\dagger_i\hat{b}^\dagger_i\hat{b}_i\hat{b}_i \rangle \nonumber\\
&=&|\phi_i(x_1)|^4 N\rho_i(N\rho_i-1),
\end{eqnarray}
where $\hat{b}_i$ annihilates the $i$-th orbital $\phi_i$ whose occupancy is $\rho_i$ [cf. Eq.~\eqref{eq:occupancy}]. As a result, the two-body correlation function [cf. Eq.~\eqref{eq:def_g2}] at $x=x_0$ can be expressed in $\rho_i$,
\begin{eqnarray}
g^{(2)}(x_0,x_0) =\frac{N\rho_i(N\rho_i-1)}{(N\rho_i)^2},
\end{eqnarray}
reproducing the result of Eq.~\eqref{eq:2cf} in the main body of the paper.

\section{Example of reentrance of superfluidity}\label{sec:append_re}

We now construct a simple toy model to explain the reentrance of superfluidity. We suppose that  in the region in question, a total number of atoms $N$ occupy $n_\text{site}=4$ lattice sites, which is the case in the simulations for $N=50$ atoms at $\eta\sqrt{N}=930$ ($2\pi\times234$kHz). As the pump rate increases, the atoms hop from the inner to the outer peaks to reduce the on-site energy cost every time $\eta$ increases across a certain pump rate $\eta^*$. For a finite interval before or after $\eta^*$, the number of atoms at each lattice site remains fixed. The number of atoms in the inner peaks before $\eta$ reaches $\eta^*$ is $n_1=\lceil N(1+8/\xi^*)/4\rceil$ [cf. Eq.~\eqref{eq:atom_num_site}], where $\xi^*=NU_s^*/\omega^2$ corresponds to $\eta^*$. We note that  $n_1$ is a fixed number and can be read out directly from the occupancy $n_1=N\rho_1=N\rho_2$ in simulations [see Section~\ref{subsec:SMI}].
When the pump rate $\eta$ increases across $\eta^*$, the state goes from $|1\rangle$ to $|2\rangle$, which, together with the other two relevant configurations, are defined as
\begin{equation}
\begin{split}
|1\rangle\equiv&|N/2-n_1,n_1,n_1,N/2-n_1\rangle \\
|2\rangle\equiv&|N/2-n_1+1,n_1-1,n_1-1,N/2-n_1+1\rangle\\
|3\rangle\equiv&|N/2-n_1+1,n_1-1,n_1,N/2-n_1\rangle\\ 
|4\rangle\equiv&|N/2-n_1,n_1,n_1-1,N/2-n_1+1\rangle.
\end{split} 
\end{equation}
In the subspace spanned by  states $\{|1\rangle,|3\rangle,|4\rangle,|2\rangle\}$, the BH Hamiltonian [Eq.~\eqref{BH_Ham}] can be written as 
\begin{eqnarray}
\hat{\mathcal{H}}_\text{BH}\approx H_0 \mathds{1}+
\begin{pmatrix}0&t&t&0\\t&\epsilon &0&t \\ t&0&\epsilon&t \\ 0&t&t&2\epsilon\end{pmatrix},
\end{eqnarray}
where $H_0=U_s[n_1^2+(N/2-n_1)^2]+\omega^2(9N/2-8n_1)$
is the energy of state $|1\rangle$ and $\epsilon=-U_s|N-4n_1+2|/2+4\omega^2$
is the energy difference between the states $|1\rangle$ and $|3\rangle$.

As $\eta$ and therefore $U_s$ increase but $\omega$ remains unchanged, $\epsilon/t$ goes from a large positive number to zero then to a large negative number. For large positive (negative) $\epsilon/t$, the lowest energy steady state is dominated by $|1\rangle$ ($|2\rangle$), and there are no correlations among the four peaks. However, for $|\epsilon/t|<10$, the ground state is a superposition of the four states and the correlations among the four peaks become large.

\section{Accuracy of the Gaussian approximation of the Wannier function}\label{sec:accuracy_BH}

The discrepancy between the simulation and analytical results stems from the inaccuracy of the Gaussian as an approximation of the Wannier function. 
Consider the on-site interaction $U_s$, whose dependence on the Wannier function is given by $\int \mathrm{d}x |W(x)|^4$ [cf. Eq.~\eqref{BH_para_us}]. To estimate the exactness of the Gaussian approximation, we calculate the values of the integral for the central peak(s)
\begin{subequations}
\begin{eqnarray}
\frac{U_{s,\text{sim}}}{g}=\dfrac{\int_{-\pi/k_c}^{\pi/k_c} \mathrm{d}x \rho^2(x)}{\int_{-\pi/k_c}^{\pi/k_c} \mathrm{d}x \rho(x)}
\end{eqnarray}
for even lattices and 
\begin{eqnarray}
\frac{U_{s,\text{sim}}}{g}=\dfrac{\int_{0}^{2\pi/k_c} \mathrm{d}x \rho^2(x)}{\int_{0}^{2\pi/k_c} \mathrm{d}x \rho(x)}
\end{eqnarray}
for odd lattices in simulations, and compare it to the prediction of the Gaussian ansatz from the BH model, which is given by
\begin{eqnarray}
\frac{U_{s,\text{Gauss}}}{g}=\sqrt{\frac{\eta k_c}{\pi}}\left(\frac{N}{2|\Delta_c|}\right)^{\frac{1}{4}}
\end{eqnarray}
\end{subequations}
in the dimensionless units. The ratio of these two values, $U_{s,\text{Gauss}}/U_{s,\text{sim}}$,
as a function of $\eta\sqrt{N}$ is shown in Fig.~\ref{wannier_obp}(d). The Gaussian always overestimates the on-site interaction, especially at low pump rates. This is consistent with the fact that the analytical results are always shifted towards the lower pump rates compared to the numerical ones in Figs.~\ref{xexpansion}, \ref{occupancy}, and \ref{wannier_obp}(a). 
However, when the pump laser is sufficiently strong, our approximation becomes close to the correct Wannier functions.

\bibliographystyle{apsrev}
\bibliography{References}
 
\end{document}